\documentclass[runningheads]{llncs}

\usepackage{amsmath}
\usepackage{amssymb}
\usepackage{graphicx}
\usepackage{fink}
\usepackage[table]{xcolor}
\usepackage[font=footnotesize]{caption}
\usepackage[vlined]{algorithm2e}
\usepackage{multirow}
\usepackage{caption}
\usepackage{subcaption}
\usepackage{float}
\usepackage[T1]{fontenc}
\usepackage{hyperref}
\usepackage {mathrsfs}

\newcommand{\muz}{$\mu Z$\xspace}
\newcommand{\ie}{{\em i.e.},\xspace}
\newcommand{\viz}{{\em viz.},\xspace}
\newcommand{\eg}{e.g.,\xspace}
\newcommand{\such}{\centerdot}

\newcommand{\algo}{\textsc{Spacer}\xspace}
\newcommand{\subheading}[1]{\noindent\textbf{#1.}}

\newcommand{\cpentry}{\emph{en}\xspace}
\newcommand{\cploop}{\emph{lp}\xspace}
\newcommand{\cperror}{\emph{er}\xspace}

\newcommand{\bigand}{\bigwedge}
\newcommand{\bigor}{\bigvee}

\newcommand {\pf} {\pi}
\newcommand{\cex}{\mathscr{C}}
\newcommand {\invar} {\mathcal{I}}

\newenvironment{pfsketch}{\paragraph{Proof Sketch.}}{\hfill\null}

\renewcommand{\implies}{\Rightarrow}

\providecommand{\SetAlgoLined}{\SetLine}
\providecommand{\DontPrintSemicolon}{\dontprintsemicolon}

\title{Automatic Abstraction in SMT-Based\\ Unbounded Software Model Checking
\thanks{
This research was sponsored by the National Science Foundation grants
no.~DMS1068829, CNS0926181 and CNS0931985, the GSRC under
contract no.~1041377, the Semiconductor Research Corporation under
contract no.~2005TJ1366, the Office of Naval Research under award
no.~N000141010188 and the CMU-Portugal Program.
This material is based upon work funded and supported by the Department of Defense
under Contract No. FA8721-05-C-0003 with Carnegie Mellon University
for the operation of the Software Engineering Institute, a federally
funded research and development center.
This material has been approved for public release and unlimited
distribution. (DM-0000279).
This is originally published by Springer-Verlag as part of the proceedings of
CAV 2013.
}
}
\titlerunning{Abstraction in SMT-based SMC}

\author{Anvesh Komuravelli \and Arie Gurfinkel \and Sagar
Chaki \and Edmund M. Clarke}
\authorrunning{A.~Komuravelli et al.}
\institute{Carnegie Mellon University, Pittsburgh,
PA, USA
}

\date{}

\begin{document}
\maketitle

\begin{abstract}
  Software model checkers based on under-approximations and SMT
  solvers are very successful at verifying safety (\ie reachability)
  properties. They combine two key ideas -- (a) \emph{concreteness}: a
  counterexample in an under-approximation is a counterexample in the
  original program as well, and (b) \emph{generalization}: a proof of safety
  of an under-approximation, produced by an SMT solver, are
  generalizable to proofs of safety of the original program. In this
  paper, we present a combination of \emph{automatic abstraction} with
  the under-approximation-driven framework. We explore two iterative
  approaches for obtaining and refining abstractions -- \emph{proof
    based} and \emph{counterexample based} -- and show how they can be
  combined into a unified algorithm. To the best of our knowledge, this is the
  first application of Proof-Based Abstraction, primarily used to verify
  hardware, to Software Verification. We have implemented a prototype
  of the framework using Z3, and evaluate it on many benchmarks from the Software
  Verification Competition. We show experimentally that our
  combination is quite effective on hard instances.
\end{abstract}


\newcommand {\unsat}{{\tt unsat}}

\section {Introduction}
\label {sec:intro}







\newcommand{\ubmc}{\textsc{2BMC}\xspace}

Algorithms based on generalizing from under-approximations are very
successful at verifying safety properties, \ie absence of bad
executions (\eg~\cite{ufo,smc_ic3,impact}). Those techniques use
what we call a \emph{Bounded Model Checking-Based Model Checking}
(\ubmc). The key idea of \ubmc is to iteratively construct an
under-approximation $U$ of the target program $P$ by unwinding its
transition relation and check whether $U$ is safe using Bounded Model
Checking (BMC)~\cite{bmc}. If $U$ is unsafe, so is $P$. Otherwise, a
proof $\pi_U$ is produced explaining
\emph{why} $U$ is safe. Finally, $\pi_U$ is generalized (if possible)
to a safety proof of $P$. Notable instances of \ubmc are based on
interpolation (\eg~\cite{ufo,impact}) or Property Directed
Reachability (PDR)~\cite{ic3,pdr} (\eg~\cite{smc_ic3,gpdr}).


At the same time, automatic abstraction refinement, such as
CounterExample Guided Abstraction Refinement (CEGAR)~\cite{cegar}, is
very effective~\cite{ufo,slam,blast}. The idea is to iteratively construct, verify, and refine
an abstraction (\ie an over-approximation) of $P$ based on
 abstract counterexamples. 
In this paper, we present \algo\footnote{Software Proof-based Abstraction with
CounterExample-based Refinement.}, an algorithm that
combines abstraction with \ubmc.

For example, consider the  safe program
$P_g$ by Gulavani et al.~\cite{gulavani} shown in Fig.~\ref{fig:abs_helps}. $P_g$ is
hard for existing \ubmc techniques. For example, $\mu Z$ engine of
Z3~\cite{z3} (v$4.3.1$) that implements Generalized PDR~\cite{gpdr}
cannot solve it within an hour. However, its abstraction $\hat{P}_g$
obtained by replacing line~7 with a non-deterministic assignment to
\texttt{t} is solved by the same engine in under a second.  Our
implementation of \algo finds a safe abstraction of $P_g$ in under a minute
(the transition relation of the abstraction we automatically computed is a
non-trivial generalization of that of $P_g$ and does not correspond to $\hat{P}_g$).

\newlength {\mywidth}
\settowidth {\mywidth} {\tt 4:     if (x>=4) {x++; y++;}}

\begin{figure}[t]
    \centering
    {\scriptsize
    \begin{minipage}{\mywidth}
    \begin{verbatim}
0: x=0; y=0; z=0; w=0;
1: while(*) {
2:   if(*) {x++; y=y+100;}
3:   else if(*) 
4:     if (x>=4) {x++; y++;}
5:   else if(y>10*w &&
             z>=100*x)
6:     {y=-y;}
7:   t=1;
8:   w=w+t; z=z+(10*t);
   }
9: assert(!(x>=4 && y<=2));
    \end{verbatim}
    \end{minipage}
    }
    \caption{A program $P_g$ by Gulavani et al.~\cite{gulavani}.}
    \label{fig:abs_helps}
    \vspace{0.1in}
\end{figure}

\begin{figure}[t]
    \centering
    \includegraphics[scale=1.2]{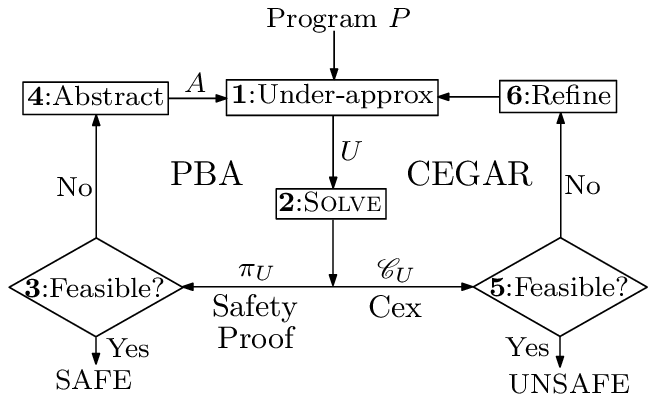}
    \caption{An overview of \algo.}
    \label{fig:flow}
\end{figure}

\newcommand{\solver}{\textsc{Solve}\xspace}


\algo tightly connects \emph{proof-based} (PBA) and
\emph{counterexample-based} (CEGAR) abstraction-refinement schemes. An
overview of \algo is shown in Fig.~\ref{fig:flow}. The input is a
program $P$ with a designated error location $\cperror$ and the output is
either SAFE with a proof that $\cperror$ is unreachable, or UNSAFE with
a counterexample to $\cperror$. \algo is sound, but obviously incomplete, \ie it is not
guaranteed to terminate.

During execution, \algo maintains an abstraction $A$ of $P$, and an
under-approximation $U$ of $A$. We require that the safety problem for
$U$ is decidable. So, $U$ is obtained by considering finitely many
finitary executions of $A$.  Initially, $A$ is any abstraction of $P$ (or $P$
itself) and $U$ is some under-approximation (step~1) of $A$.  In each
iteration, the main decision engine, called \solver, takes $U$ and
outputs either a proof $\pi_U$ of safety (as an inductive invariant)
or a counterexample trace $\cex_U$ of $U$ (step~2). In practice, \solver
is implemented by an interpolating SMT-solver
(\eg~\cite{mathsat,itp+pa}), or a generalized Horn Clause solver
(\eg~\cite{duality,hsfc,gpdr}).  If $U$ is safe and $\pi_U$ is also valid for
$P$ (step~3), \algo terminates with SAFE; otherwise, it constructs a
new abstraction $\hat{A}$ (step~4) using $\pi_U$, picks an
under-approximation $\hat{U}$ of $\hat{A}$ (step~1), and goes into the next
iteration. If $U$ is unsafe and $\cex_U$ is a feasible trace of $P$
(step~5), \algo terminates with UNSAFE; otherwise, it refines the
under-approximation $U$ to refute $\cex_U$ (step~6) and goes to the next
iteration. \algo is described in Section~\ref{sec:details} and a detailed
run of the algorithm on an example is given in Section~\ref{sec:overview}.

Note that the left iteration of \algo (steps 1, 2, 3, 4) is PBA: in
each iteration, an under-approximation is solved, a new abstraction based
on the proof is computed and a new under-approximation is
constructed. To the best of our knowledge, this is the first application of PBA to
Software Model Checking. The right iteration (steps 1, 2, 5, 6) is
CEGAR: in each iteration, (an under-approximation of) an abstraction is
solved and refined by eliminating spurious counterexamples. \algo
exploits the natural duality between the two.

While \algo is not complete, each iteration makes progress either by
proving safety of a bigger under-approximation, or by refuting a
spurious counterexample. Thus, when resources are exhausted, \algo can
provide useful information for other verification attempts and for
increasing confidence in the program.


We have implemented \algo using \muz~\cite{gpdr} as \solver (Section~\ref{sec:impl}) and evaluated it on
many benchmarks from the 2nd Software Verification
Competition\footnote{\url{http://sv-comp.sosy-lab.org}}
(SV-COMP'13). Our experimental results (see
Section~\ref{sec:results}) show that the
combination of \ubmc and abstraction outperforms \ubmc on hard benchmarks.

In summary, the paper makes the following contributions: (a) an
algorithm, \algo, that combines abstraction and \ubmc and tightly
connects proof- and counterexample-based abstractions, (b) an
implementation of \algo using \muz engine of Z3 and (c) experimental
results showing the effectiveness of \algo.




\section{Overview}
\label{sec:overview}
\begin{figure}[t]
\begin {subfigure}[b]{.5\textwidth}
{\scriptsize
\begin{verbatim}
 0: x=0; y=0; z=0; w=0; c=nd();
 1: b=nd();
 2: while (0<c<=b) {
 3:   if (nd()) {x++; y=y+100;}
 4:   else if (nd() && x>=4) {x++; y++;}
 5:   else if (y>10*w && z>=100*x) {y=-y;}
 6:   else assume (0);
 7:   w++; z=z+10; c--; 
    }
 8: assert (!(c==0 && x>=4 && y<=2));
\end{verbatim}}
\caption {}
\label {fig:P}
\end{subfigure}
\begin {subfigure}[b]{.5\textwidth}
\includegraphics[scale=1]{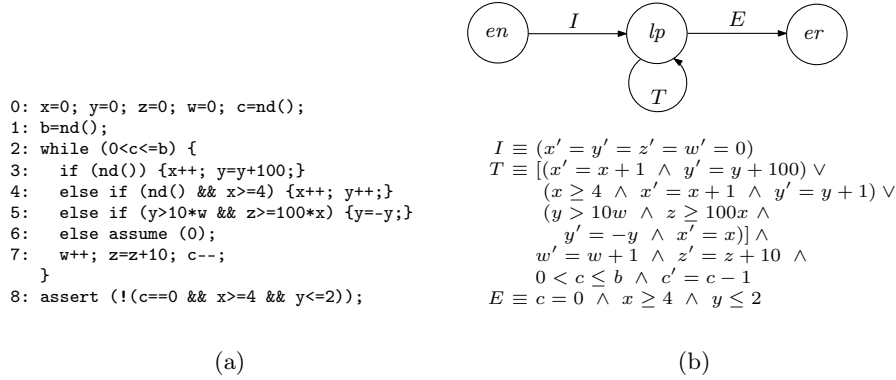}
\[\scriptsize
\begin{array}{rcl}
  I & \equiv & (x'=y'=z'=w'=0) \\
  T & \equiv & [(x'=x+1 ~\wedge~ y'=y+100) \vee{}\\
    &        & ~(x \geq 4 ~\wedge~ x'=x+1 ~\wedge~ y'=y+1) \vee{}\\
    &        & ~(y > 10w ~\wedge~ z\geq 100x ~\wedge \\
    &        & ~\quad y'=-y ~\wedge~ x'=x)] \wedge{}\\
    &        & w'=w+1 ~\wedge~ z'=z+10 ~\wedge{}\\
    &        &  0<c\leq\emph{b} ~\wedge~ c'=c-1\\
  E & \equiv & c=0 ~\wedge~ x \geq 4 ~\wedge~ y \leq 2
\end{array}
\]
\caption {}
\label {fig:lbe}
\end{subfigure}
\caption {(a) A program $P$ and (b) its transition system.}
\label {fig:overview_orig}
\end{figure}

In this section, we illustrate \algo on the program $P$ shown in
Fig.~\ref{fig:P}.  Function \texttt{nd()} returns a value
non-deterministically and \texttt{assume(0)} aborts an execution.
Thus, at least one of the updates on lines~3,~4 and~5
must take place in every iteration of the loop on line~2.  Note that
the variable \texttt{c} counts down the number of iterations of the loop to $0$, upper
bounded by \texttt{b}. A restriction to \texttt{b} is an
under-approximation of $P$. For example, adding `\texttt{assume(b<=0);}'
to line~1 corresponds to the under-approximation of $P$ that allows only
loop-free executions; adding `\texttt{assume(b<=1);}' to line~1 corresponds
to the under-approximation that allows at most one execution through the loop,
etc. While in this example the \emph{counter variable} \texttt{c} is
part of $P$, we synthesize such variables automatically in practice
(see~Section~\ref{sec:impl}).

Semantically, $P$ is given by the transition system shown in
Fig~\ref{fig:lbe}. The control locations \cpentry, \cploop, and \cperror
correspond to lines~0,~2, and~8 in $P$, respectively. An edge from
$\ell_1$ to $\ell_2$ corresponds to all loop-free executions starting at
$\ell_1$ and ending at $\ell_2$. For example, the self-loop on \cploop
corresponds to the body of the loop. Finally, every edge is labeled by
a formula over current (unprimed) and next-state (primed) variables
denoting the semantics of the corresponding executions. Hence, $I$ and
$E$ denote the initial and error conditions, respectively, and $T$
denotes the loop body. In the rest of the paper, we do not distinguish
between semantic and syntactic representations of programs.

Our goal is to find a safety proof for $P$, \ie a labeling $\pf$ of
\cpentry, \cploop and \cperror with a set of formulas (called \emph{lemmas}) that satisfies
safety, initiation and inductiveness:
\begin{align*}
  \bigwedge \pf(\cperror) &\Rightarrow \bot, &
  \top &\Rightarrow \bigwedge \pf(\cpentry), & 
  \forall \ell_1,\ell_2 \such \left(\bigwedge \pf(\ell_1) ~\land~ \tau(\ell_1,
\ell_2)\right) &\Rightarrow
\bigwedge \pf(\ell_2)'.
\end{align*}
where $\tau(\ell_1, \ell_2)$ is the label of edge from $\ell_1$ to $\ell_2$, and
for an expression $X$, $X'$ is obtained from $X$ by priming all variables.
In the following, we refer to Fig.~\ref{fig:flow} for the steps of the algorithm. 


\begin{figure}[t]
\begin{subfigure}[c]{.5\textwidth}
{\scriptsize \[
\begin{array}{rcl}
  \hat{I}_1 & \equiv & (x'=y'=z'=w'=0)\\
  \hat{T}_1 & \equiv & [(x'=x+1) \vee{}\\
           &        & ~(x \geq 4 ~\wedge~ x'=x+1) \vee{}\\
           &        & ~(y > 10w ~\wedge~ z\geq 100x)] \wedge{}\\
           &        &  0<c\le b ~\wedge~ c'=c-1\\
  \hat{E}_1 & \equiv & c=0 ~\land~ x \geq 4
\end{array}
\]}
\caption {$\hat{P}_1$}
\label {fig:abs1}
\end{subfigure}
\begin{subfigure}[c]{.5\textwidth}
{\scriptsize \[
\begin{array}{rcl}
  \hat{I}_2 & \equiv&  (x'=y'=z'=w'=0)\\
  \hat{T}_2 & \equiv & [(x'=x+1 ~\wedge~ y'=y+100) \vee{}\\
           &        & ~(x \geq 4 ~\wedge~ x'=x+1 ~\wedge~ y'=y+1) \vee{}\\
           &        & ~(y > 10w ~\wedge~ z\geq 100x)] \wedge{}\\
           &        &  0<c\le b ~\wedge~ c'=c-1\\
  \hat{E}_2 & \equiv & c=0 ~\land~ x \geq 4 ~\wedge~ y \leq 2
\end{array}
\]}
\caption {$\hat{P}_2$}
\label {fig:abs2}
\end{subfigure}
\caption {Abstractions $\hat{P}_1$ and $\hat{P}_2$ of $P$ in Fig.~\ref{fig:lbe}.}
\label {fig:overview_abs}
\end{figure}


\subheading{Steps 1 and 2} Let $U_1$ be
the under-approximation obtained from $P$ by conjoining $(b \le 2)$ to
$T$. It is safe, and suppose that \solver returns the safety proof $\pf_1$,
shown in Fig.~\ref{fig:lemmas}.

\subheading{Step 3} To check whether $\pf_1$ is also a safety
proof of the concrete program $P$, we extract a \emph{Maximal
Inductive Subset} (MIS), $\invar_1$ (shown in Fig.~\ref{fig:invs}), of
$\pf_1$, with respect to $P$. That is, for every location $\ell$, $\invar_1(\ell)
\subseteq \pf_1(\ell)$, and $\invar_1$ satisfies the initiation and
inductiveness conditions above. $\invar_1$ is an inductive invariant of
$P$, but is not safe (\cperror\ is not labeled with $\bot$). Hence,
$\pf_1$ does not contain a feasible proof, and another iteration
of \algo is required.



\begin{figure}[t]
  \newcommand{\spc}{\phantom{\{}}
\begin{subfigure}[b]{.3\textwidth}
\[
{\scriptsize
\begin{array}{lcl}
  \cpentry & : & \{\}\\
  \cploop & : & \{(z \leq 100x - 90 \vee{}\\
          &   & \multicolumn{1}{r}{y \leq 10w),} \\
          &   & \spc z \leq 100 x, x \leq 2 \\
          &   & \spc (x \leq 0 \vee c \leq 1)\\
          &   & \spc (x \leq 1 \vee c \leq 0)\}\\
\cperror  & : & \{ \bot \}
\end{array}}
\]
\caption {$\pf_1$: safety proof of $U_1$.}
\label {fig:lemmas}
\end{subfigure}
\begin{subfigure}[b]{.3\textwidth}
\[
{\scriptsize
\begin{array}{lcl}
  \cpentry & : & \{\}\\
  \cploop & : & \{(z \leq 100x - 90 \vee{}\\
          &   & \multicolumn{1}{r}{y \leq 10w),} \\
          &   & \spc z \leq 100 x \}\\
          &   & \\
          &   & \\
\cperror  & : & \{ \}
\end{array}}
\]
\caption {$\invar_1$: invariants of $P$.}
\label {fig:invs}
\end{subfigure}
\begin{subfigure}[b]{.3\textwidth}
\[
{\scriptsize
\begin{array}{lcl}
  \cpentry & : & \{\}\\
  \cploop & : & \{(z \leq 100x - 90 \vee{}\\
          &   & \multicolumn{1}{r}{y \leq 10w),} \\
          &   & \spc z \leq 100 x, y\geq 0, \\
          &   & \spc (x \leq 0 \vee y \geq 100) \}\\
          &   & \\
\cperror  & : & \{ \bot \}
\end{array}}
\]
\caption {$\pf_3$: safety proof of $U_3$.}
\label {fig:proof}
\end{subfigure}

\caption {Proofs and invariants for the running example in Section~\ref{sec:overview}.}
\label {fig:overview_pf}
\end{figure}

\subheading{Step 4} We obtain an abstraction $\hat{P}_1$ of $P$
for which, assuming the invariants in $\invar_1$,
$\pf_1$ is a safety proof for the first two iterations of the loop (\ie when $b \leq
2$). For this example, let $\hat{P}_1$ be as shown in Fig.~\ref{fig:abs2}.
Note that $\hat{T}_1$ has no constraints on the next-state values of $z$, $y$ and
$w$. This is okay for $\pf_1$ as $\invar_1(\cploop)$ already
captures the necessary relation between these variables. In other words,
while $\hat{T}_1$ is a structural (or \emph{syntactic})
abstraction~\cite{babic07}, we consider its restriction to the invariants
$\invar_1$ making it a more expressive, \emph{semantic} abstraction.
The next iteration of \algo is described below.

\subheading{Steps 1 and 2} Let $U_2$ be the under-approximation
obtained from $\hat{P}_1$ by conjoining $(b \le 4) \land \invar_1 \land
\invar'_1$ to $\hat{T}_1$. It is not safe and let \solver return a counterexample $\cex_2$ as
the pair $\langle \bar{\ell}, \bar{s} \rangle$ of the following
sequences of locations and  states, corresponding to incrementing $x$ from $0$
to $4$ with an unconstrained $y$:
\begin{equation}\small
\begin{aligned}
\bar{\ell} &\equiv \langle \cpentry,\cploop,\cploop,\cploop,\cploop,
			\cploop,\cperror \rangle \\
\bar{s} &\equiv \langle (0,0,0,0,0,0), (0,0,0,0,4,4), (1,0,0,0,3,4), (2,0,0,0,2,4),\\ &\phantom{{}\equiv{}\langle}
 (3,0,0,0,1,4), (4,3,0,0,0,4), (4,3,0,0,0,4) \rangle
\end{aligned}
\label{eq:cex}
\end{equation}
where a state is a valuation to the tuple $(x,y,z,w,c,b)$.

\subheading{Steps 5 and 6} $\cex_2$ is infeasible in $P$ as the last state
does not satisfy $E$. $\hat{P}_1$ is refined to $\hat{P}_2$, say as shown in
Fig.~\ref{fig:abs2}, by adding the missing constraints on $y$.

\subheading{Steps 1 and 2} Let $U_3$
be the under-approximation obtained from $\hat{P}_2$ by conjoining $(b \le 4)
\land \invar_1 \land \invar'_1$ to $\hat{T}_2$.
It is safe, and let \solver\ return the proof
$\pf_3$ shown in Fig.~\ref{fig:proof}.

\subheading{Step 3} $\pf_3$ is a MIS of itself, with
respect to $P$. Thus, it is a safety proof for $P$ and \algo
terminates.

While we have carefully chosen the under-approximations to save space,
the abstractions, lemmas and invariants shown above were all computed
automatically by our prototype implementation starting with the initial
under-approximation of $b \le 0$ and incrementing the upper bound by $1$, each
iteration. Even on this small example, our prototype, built using $\mu Z$, is
five times faster than $\mu Z$ by itself.




\newcommand {\power} [1] {2^{#1}}
\newcommand {\bexpr}[1] {\text{BExpr}({#1})}
\newcommand {\partialto} {\rightharpoonup}
\newcommand {\under} {\preceq}
\newcommand {\sunder} {\prec}
\newcommand {\id} {\textnormal{{\em id}}}

\newcommand {\triple} [3] {$\langle${#1}$,$ {#2}$,$ {#3}$\rangle$}
\newcommand {\quadruple} [4] {$\langle${#1}$,$ {#2}$,$ {#3}$,$ {#4}$\rangle$}
\newcommand {\quintuple} [5] {$\langle${#1}$,$ {#2}$,$ {#3}$,$ {#4}$,$ {#5}$\rangle$}

\newcommand {\prog} {\quintuple {$L$} {$\ell^o$} {$\ell^e$} {$V$} {$\tau$} }
\newcommand {\vprog} [1] {\quintuple {$L_{#1}$} {$\ell^o_{#1}$} {$\ell^e_{#1}$}
                                     {$V_{#1}$} {$\tau_{#1}$}}
\newcommand {\aprog} [1] {\quintuple {$L_{#1}$} {$\ell^o_{#1}$} {$\ell^e_{#1}$}
                                     {$V_{#1}$} {$\hat{\tau}_{#1}$}}

\section {Preliminaries}
\label {sec:prelims}

This section defines the terms and notation used in the rest of the
paper.

\begin{definition}[Program]
  A program $P$ is a tuple \prog where
\begin {enumerate}
\item $L$ is the set of control locations,
\item $\ell^o \in L$ and $\ell^e \in L$ are the unique initial and
  error locations,
\item $V$ is the set of all program variables $($Boolean or Rational$)$, and
\item $\tau : L \times L \to \bexpr{V \cup V'}$ is a map from pairs
  of locations to Boolean expressions over $V \cup V'$ in propositional
  Linear Rational Arithmetic.
\end{enumerate}
\end{definition}

Intuitively, $\tau(\ell_i, \ell_j)$ is the
relation between the current values of $V$ at $\ell_i$ and the next
values of $V$ at $\ell_j$ on a transition from $\ell_i$ to
$\ell_j$. We refer to $\tau$ as the transition relation.
Without loss of generality, we assume that
%
$\forall \ell \in L \such \tau(\ell, \ell^o) = \bot \land \tau(\ell^e,\ell) = \bot$.
We refer to the components of $P$ by a subscript, \eg $L_P$.  

Fig.~\ref{fig:lbe} shows an example program with $L = \{\cpentry,
\cploop, \cperror\}$, $\ell^o = \cpentry$, $\ell^e = \cperror$, $V =
\{x,y,z,w,c,b\}$, $\tau(\cpentry,\cploop) = I$, $\tau(\cploop,
\cploop) = T$, $\tau(\cploop, \cperror) = E$.

%


Let $P = $\prog be a program. A {\em control path} of $P$ is a
finite\footnote{In this paper, we deal with safety properties only.}
sequence of control locations \triple {$\ell^o=\ell_0$} {$\ell_1,
  \dots$} {$\ell_k$}, beginning with the initial location $\ell^o$,
such that $\tau(\ell_i,\ell_{i+1}) \neq \bot$ for $0 \le i < k$.
A {\em state} of $P$ is a valuation to all the variables
in $V$. A control path \triple {$\ell^o=\ell_0$} {$\ell_1, \dots$}
{$\ell_k$} is called {\em feasible} iff there is a sequence of states
\triple {$s_0$} {$s_1, \dots$} {$s_k$} such that
\begin{equation} \label{eq:sat}
\forall 0 \leq i < k \such \tau (\ell_i, \ell_{i+1}) [V \leftarrow
  s_i, V' \leftarrow s_{i+1}] = \top
\end{equation}
\ie each successive and corresponding pair of locations and states satisfy~$\tau$.

For example, \quadruple{\cpentry}{\cploop}{\cploop}{\cploop} is a
feasible control path of the program in Fig.~\ref{fig:lbe} as the
sequence of states \quadruple {$(0,0,0,0,0,0)$} {$(0,0,0,0,2,2)$}
{$(1,100,1,10,$ $1,2)$} {$(2,200,2,20,0,2)$} satisfies~\eqref{eq:sat}.


A location $\ell$ is \emph{reachable} iff there exists a feasible control
path ending with $\ell$.  $P$ is \emph{safe} iff $\ell^e$ is
\emph{not} reachable.
For example, the program in Fig.~\ref{fig:lbe} is safe.  $P$ is
\emph{decidable}, when the safety problem of $P$ is decidable. For
example, the program $U$ obtained from $P$ in Fig.~\ref{fig:lbe} by
replacing $b$ with $5$ is decidable because (a)~$U$ has finitely
many feasible control paths, each of finite length and (b) Linear
Arithmetic is decidable.

%

\begin{definition}[Safety Proof]
\label{def:proof}
A \emph{safety proof} for $P$ is a map $\pf : L \to
\power{\bexpr{V}}$ such that $\pf$ is safe and inductive, \ie
\begin{align*}
  \bigand \pf(\ell^e) &\Rightarrow \bot, & 
  \top &\Rightarrow \bigand \pf (\ell^o), &
\forall \ell_i, \ell_j \in L \such
\left(\bigand \pf(\ell_i) ~\land~ \tau(\ell_i,\ell_j)\right)
  &\Rightarrow \bigand \pf(\ell_j)'.
\end{align*}
\end{definition}

For example, Fig.~\ref{fig:proof} shows a safety proof for the program
in Fig.~\ref{fig:lbe}. Note that whenever $P$ has a safety proof, $P$
is safe. 

A \emph{counterexample to safety} is a pair $\langle \bar{\ell}, \bar{s} \rangle$ such that
$\bar{\ell}$ is a feasible control path in $P$ ending with $\ell^e$ and
$\bar{s}$ is a corresponding sequence of states satisfying $\tau$ along
$\bar{\ell}$. For example, $\hat{P}_2$ in Fig.~\ref{fig:abs2} admits the
counterexample $\cex_2$ shown in \eqref{eq:cex} in Section~\ref{sec:overview}.

\begin{definition}[Abstraction Relation]
\label{def:abs_rel}
Given two programs, $P_1 =$ \vprog{1} and $P_2 =$ \vprog{2}, $P_2$ is
an \emph{abstraction} $($\ie an over-approximation$)$ of $P_1$ via a surjection
$\sigma : L_1 \to L_2$, denoted $P_1 \under_{\sigma} P_2$, iff
\begin{align*}
V_1 &= V_2, & 
\sigma (\ell^o_1) &= \ell^o_2, &
\sigma (\ell^e_1) &= \ell^e_2, &
\forall \ell_i, \ell_j \in L_1 \such \tau_1(\ell_i,\ell_j) &\Rightarrow
  \tau_2 (\sigma(\ell_i), \sigma(\ell_j)).
\end{align*}
%
$P_1$ is called a \emph{refinement} $($\ie an under-approximation$)$ of $P_2$.
We say that $P_2$ strictly abstracts $P_1$ via $\sigma$, denoted $P_1
\sunder_\sigma P_2$, iff $(P_1 \under_\sigma P_2) \land \neg \exists
\nu \such (P_2 \under_\nu P_1)$. When $\sigma$ is not important, we drop the
subscript.
\end{definition}

That is, $P_2$ abstracts $P_1$ iff there is a surjective map $\sigma$ from
$L_1$ to $L_2$ such that every feasible transition of $P_1$
corresponds (via $\sigma$) to a feasible transition of $P_2$. For example, if $P_1$
is a finite unrolling of $P_2$, then $\sigma$ maps the locations of $P_1$ to the
corresponding ones in $P_2$. $P_2$
\emph{strictly abstracts} $P_1$ iff $P_1 \under P_2$ and there is no
surjection $\nu$ for which $P_2 \under_\nu P_1$. For example, $P
\sunder_\id \hat{P}_1$, where $P$ is in Fig.~\ref{fig:lbe} and
$\hat{P}_1$ is in Fig.~\ref{fig:abs1}.
%

We extend $\sigma: L_1 \to L_2$ from locations to control paths in the
straightforward way. For a counterexample $\cex = \langle \bar{\ell}, \bar{s}
\rangle$, we define $\sigma(\cex) \equiv \langle \sigma(\bar{\ell}), \bar{s}
\rangle$. For a transition relation $\tau$ on $L_2$, we
write $\sigma(\tau)$ to denote an embedding of $\tau$ via $\sigma$,
defined as follows: $\sigma(\tau) (\ell_1,\ell_2) = \tau
(\sigma(\ell_1), \sigma(\ell_2))$. For example, in the definition above, if
$P_1 \under_\sigma P_2$, then $\tau_1 \implies \sigma(\tau_2)$.


\newcommand {\lemmas} {\mathcal{L}}
\newcommand {\res} {\text{{\em result}}}
\newcommand {\new} {\text{{\em new}}}
\newcommand {\feas} {\textnormal{\emph{feas}}}
\newcommand {\R} {\mathcal{R}}
\newcommand {\cpaths}[1] {\text{{\em ctrl\_paths}}({#1})}
\newcommand {\conc}[1] {\dot{#1}}

\section {The Algorithm}
\label {sec:details}

\begin{figure}[t]
    \begin{subfigure}{.5\textwidth}
    \begin{algorithm}[H]
    \scriptsize
    \DontPrintSemicolon
    \SetKwData{Global}{{\bf global}}
    \SetKwData{Assume}{{\bf requires}}
 
    \Global{$P : \text{prog}$}\;
    \Global{$\invar : L_P \to \power{\bexpr{V_P}}$}\;
    \;
    \algo$(\;)$\;
    \Begin{
\nl	$A$ := $P$, $\invar := \emptyset$ \;
\nl	$(U,\sigma)$ := \textsc{InitU}$(A)$\;
\nl	\While {{\tt true}} {
\nl                $(\res,\pf,\cex)$ := \solver$(U_\invar)$\;
\nl                \If {\res\ is {\sc Safe}} {
\nl                        $\invar = \invar \cup \textsc{ExtractInvs}(A,U,\pf)$\;
\nl                        \If {$\bigand \invar(\ell^e_P) \implies \bot$} {
\nl                                \Return {\sc Safe}
                           }
\nl                        $(A, U)$ := {\sc Abstract}$(A,U,\pf)$\;
\nl                        \mbox{$(U,\sigma)$ := {\sc NextU}$(A, U)$} \;
                   } \Else {
\nl                        \mbox{$(\feas,A,U)$ := {\sc Refine}$(A,U,\cex)$\;}
\nl                        \If {$\feas$} {
\nl                                \Return {\sc Unsafe}
                        }
                }
        }
    }
    \;
    {\sc Adapt}$(U:\text{prog},\tau:\text{trans},\sigma:L_U
    \to L_P)$\;
    ~~\Assume{$\tau$ : transition relation on $L_P$}\;
    \Begin {
\nl        \Return $U[\tau_U \leftarrow \left( \tau_U \land \sigma(\tau) \right)]$
    }
    \;
    {\sc NextU}$(A:\text{prog}, U:\text{prog})$\;
    ~~\Assume{$U \under_{\sigma} A$}\;
    \Begin {
\nl    	\Return $(\hat{U},\sigma_2)$ s.t. $U \sunder_{\sigma_1} \hat{U} \under_{\sigma_2} A$,\;
      $~~\sigma = \sigma_2 \circ \sigma_1$ and\;
      $~~\textsc{Adapt}(U, \tau_P, \sigma) \sunder
                 \textsc{Adapt}(\hat{U}, \tau_P, \sigma_2)$
    }
    \end{algorithm}
    \end{subfigure}
    \begin{subfigure}{.5\textwidth}
    \begin{algorithm}[H]
    \scriptsize
    \DontPrintSemicolon
    \setcounter{AlgoLine}{15}
   \SetKwData{Assume}{{\bf requires}}
    {\sc Abstract}$(A,U:\text{prog},\pf:\text{proof of $U$})$\;
    ~~\Assume{$U \under_\sigma A, \tau_U = \sigma(\tau_A) \land \rho$}\;
    \Begin {
\nl	   let $\hat{U}$ be s.t. $L_{\hat{U}} = L_U$,\;
       ~~$\tau_{\hat{U}} \equiv \sigma(\hat{\tau}_P) \land \hat{\rho}$ with $\tau_P
           \Rightarrow \hat{\tau}_P$,\; ~~$\rho \Rightarrow \hat{\rho}$, and $\pf$
is a safety proof of  $\hat{U}_\invar$\;
\nl	   \Return $(A[\tau_A \leftarrow \hat{\tau}_P], \hat{U})$
    }
    \;
    {\sc Refine}$(\hat{A},\hat{U}:\text{prog},\cex:\text{cex of $\hat{U}$})$\;
    ~~\Assume{$\hat{U} \under_\sigma \hat{A}$}\;
    \Begin {
\nl        $\feas$ := {\sc IsFeasible}$(\sigma(\cex),P)$\;
\nl        \If {$\neg\feas$} {
\nl                let $A \sunder_\id \hat{A}$
	s.t.	$\neg${\sc IsFeasible}\makebox[0pt][l]{$(\sigma(\cex),A_\invar)$}\;
\nl                $U$ := {\sc Adapt}$(\hat{U},\tau_A,\sigma)$\;
\nl                \Return ({\tt false}, $A$, $U$)
        }
\nl        \Return ({\tt true}, None, None)
    }
    \;
    {\sc ExtractInvs}$(A,U:\text{prog},\pi:\text{proof of $U$})$\;
    ~~\Assume {$U \under_\sigma A$}\;
    \Begin {
\nl        $\R : L_P \to \power{\bexpr{V_P}} := \emptyset$\;
\nl        \For {$\ell \in L_U$} {
\nl                add $\bigand \pi(\ell)$ to $\R(\sigma(\ell))$\;
        }
\nl        \For {$\ell \in L_P$} {
	\nl                $\R(\ell)$ := {\em conjuncts}$(\bigor \R(\ell))$
        }
\nl        \While {$\exists \ell_i, \ell_j \in L_P, \varphi \in \R(\ell_j)$ s.t.\;
                ~~~~~~$\neg \left( \R(\ell_i) \land
                                \invar(\ell_i) \land
                                \tau_P(\ell_i, \ell_j) \Rightarrow \varphi'
                                \right)$} {
\nl                $\R(\ell_j) \text{ := } \R(\ell_j) \setminus \{\varphi\}$
        }
\nl         \Return $\R$
    }
    \end{algorithm}
    \end{subfigure}
\caption{Pseudo-code of \algo.}
\label{fig:code}
\end{figure}

In this section, we describe \algo at a high-level. Low-level details
of our implementation are described in Section~\ref{sec:impl}. The
pseudo-code of \algo is shown in Fig.~\ref{fig:code}. The top level
routine \algo decides whether an input program $P$ (passed through the
global variable) is safe. It maintains (a) invariants $\invar$ such
that $\invar(\ell)$ is a set of constraints satisfied by all the
reachable states at location $\ell$ of $P$ (b) an abstraction $A$ of
$P$, (c) a decidable under-approximation $U$ of $A$ and (d) a
surjection $\sigma$ such that $U \under_\sigma A$.  \algo ensures that
$P \under_\id A$, \ie $A$ differs from $P$ only in its transition
relation.  Let $A_\invar$ denote the restriction of $A$ to the
invariants in $\invar$ by strengthening $\tau_A$ to $\lambda \ell_1,
\ell_2 \such \invar(\ell_1) \land \tau_A (\ell_1, \ell_2) \land
\invar(\ell_2)'$. Similarly, let $U_\invar$ denote the strengthening
of $\tau_U$ to $\lambda \ell_1, \ell_2 \such \invar(\sigma(\ell_1)) \land \tau_U (\ell_1, \ell_2)
\land \invar(\sigma(\ell_2))'$.
\algo assumes the existence of an oracle, \solver, that decides
whether $U_\invar$ is safe and
returns either a safety proof or a counterexample.




\algo initializes $A$ to $P$ and $\invar$ to the empty map (line~1), calls $\textsc{InitU}(A)$ to initialize $U$ and
$\sigma$ (line~2) and enters the main loop (line~3). In each iteration,
safety of $U_\invar$ is checked with \solver (line~4).  If $U_\invar$ is safe, the
safety proof $\pf$ is checked for feasibility w.r.t. the original
program~$P$, as follows. First, $\pf$ is mined for new invariants of
$P$ using {\sc ExtractInvs} (line~6). Then, if the invariants at
$\ell^e_P$ are unsatisfiable (line~7), the error location is
unreachable and \algo returns {\sc Safe} (line~8).  Otherwise, $A$ is
updated to a new proof-based abstraction via {\sc Abstract} (line~9),
and a new under-approximation is constructed using \mbox{{\sc NextU}}
(line~10).  If, on the other hand, $U_\invar$ is unsafe at line~4, the counterexample $\cex$ is validated
using {\sc Refine} (line~11). If $\cex$ is feasible, \algo returns {\sc
  Unsafe} (line~13), otherwise, both $A$ and $U$ are refined (lines~20 and~21).

Next, we describe these routines in detail. Throughout, fix $U$, $\sigma$ and
$A$ such that $U \under_\sigma A$.

\subheading{\textsc{ExtractInvs}} For every $\ell \in L$, the lemmas
of all locations in $L_U$ which map to $\ell$, via the surjection
$\sigma : L_U \to L_P (=L_A)$, are first collected into $\R(\ell)$
(lines~25--26). The disjunction of $\R(\ell)$ is then broken down into
conjuncts and stored back in $\R(\ell)$ (lines~27--28). For e.g., if $\R(\ell) = \{\phi_1,
\phi_2\}$, obtain $\phi_1 \lor \phi_2 \equiv \bigand_j \psi_j$ and
update $\R(\ell)$ to $\{\psi_j\}_j$.
Then, the invariants are extracted as the maximal subset of $\R(\ell)$
that is mutually inductive, relative to $\invar$, w.r.t. the concrete transition relation $\tau_P$.
This step uses the iterative algorithm on lines~29--30 and is similar to
\textsc{Houdini}~\cite{houdini}.

\noindent\textbf{\textsc{Abstract}} 
first constructs an abstraction $\hat{U}$ of $U$, such that $\pf$ is a safety
proof for $\hat{U}_\invar$ and then, uses the transition relation of $\hat{U}$ to get the new
abstraction.  W.l.o.g., assume that~$\tau_U$ is of the form
$\sigma(\tau_A) \land \rho$.  That is, $\tau_U$ is an embedding of
$\tau_A$ via $\sigma$ strengthened with $\rho$. An abstraction $\hat{U}$ of $U$
is constructed such that $\tau_{\hat{U}} = \sigma(\hat{\tau}_P) \land \hat{\rho}$,
where $\hat{\tau}_P$ abstracts the concrete transition relation
$\tau_P$,  $\hat{\rho}$ abstracts $\rho$ and $\pf$ proves $\hat{U}_\invar$ (line~16). The new abstraction
is then obtained from $A$ by replacing the transition relation by
$\hat{\tau}_P$ (line~17).

\noindent\textbf{\textsc{NextU}} returns the next under-approximation $\hat{U}$ to be
solved. It ensures that $U \sunder \hat{U}$ (line~15), and that the
surjections between $U$, $\hat{U}$ and $A$ compose so that the
corresponding transitions in $U$ and $\hat{U}$ map to the same
transitions of the common abstraction $A$. Furthermore, to
ensure progress, {\sc NextU} ensures that $\hat{U}$ contains {\em more
  concrete} behaviors than $U$ (the last condition on
line~15). The helper routine \textsc{Adapt} strengthens the transition relation
of an under-approximation by an embedding (line~14). 

\noindent\textbf{\textsc{Refine}} checks if the counterexample $\cex$, via
$\sigma$, is feasible in the
original program $P$ using {\sc IsFeasible} (line~18). If $\cex$ is
feasible, {\sc Refine} returns saying so (line~23). Otherwise, $\hat{A}$ is
refined to $A$ to (at least) eliminate $\cex$ (line~20). Thus,
$A \sunder_\id \hat{A}$. Finally, $\hat{U}$ is strengthened with the refined
transition relation via \mbox{{\sc Adapt} (line~21).}

The following statements show that \algo is sound and maintains progress.
The proofs of the statements are included in the appendix.

\begin{lemma}[Inductive Invariants]
  \label{lem:ind_invs}
  In every iteration of \algo, $\invar$ is inductive with respect to $\tau_P$.
\end{lemma}

\begin{theorem}[Soundness]
\label{thm:soundness}
  $P$ is safe $($unsafe$)$ if \algo returns {\sc Safe} $(${\sc
    Unsafe}$)$.
\end{theorem}



\begin{theorem}[Progress]
\label{thm:progress}
Let $A_i$, $U_i$, and $\cex_i$ be the values of $A$, $U$, and $\cex$
in the $i^{th}$ iteration of \algo with $U_i \under_{\sigma_i} A_i$ and let
$\conc{U_i}$ denote the concretization of $U_i$, \ie result of
{\sc Adapt}$(U_i,\tau_P,\sigma_i)$. Then, if $U_{i+1}$ exists,
\begin{enumerate}
\item if $U_i$ is safe, $U_{i+1}$ has strictly more concrete behaviors, \ie
$\conc{U}_i \sunder \conc{U}_{i+1}$,
\item if $U_i$ is unsafe, $U_{i+1}$ has the same concrete behaviors, \ie
    $\conc{U}_i \under_\id \conc{U}_{i+1}$ and $\conc{U}_{i+1} \under_\id
    \conc{U}_i$, and
\item if $U_i$ is unsafe, $\cex_i$ does not repeat in future, \ie $\forall j>i
\such \sigma_j(\cex_j) \neq \sigma_i (\cex_i)$.
\end{enumerate}
\end{theorem}

In this section, we presented the high-level structure of \algo.  Many
routines (\textsc{InitU}, \textsc{ExtractInvs},
\textsc{Abstract}, \textsc{NextU}, \textsc{Refine},
\textsc{IsFeasible}) are only presented by their interfaces with
their implementation left open. In the next section, we complete the picture
by describing the implementation used in our prototype.


\newcommand {\wto} {<}
\newcommand {\wwto} {\le}
\newcommand {\hds}[1] {\text{{\em hds}}({#1})}
\newcommand {\ctr} {\text{{\em ctr}}}
\newcommand {\bd} {\text{{\em bound}}}
\newcommand {\nat} {\mathbb{N}}
\newcommand {\bp}[1] {\tilde{#1}}
\newcommand {\last}[1] {\text{{\em last}}({#1})}
\newcommand {\ctx}{\text{{\em ctx}}}
\newcommand {\postlem}[1]{\text{{\em post\_lemma}}({#1})}
\newcommand {\assum} {\Sigma}
\newcommand {\bvals} {\text{{\em bvals}}}
\newcommand {\core}[1] {\text{{\em core}}({#1})}
\newcommand {\ltag}[1] {N_{#1}}
\newcommand {\rtag}[2] {E_{{#1},{#2}}}
\newcommand {\children}[1] {\text{{\em children}}({#1})}
\newcommand {\parents}[1] {\text{{\em parents}}({#1})}

\newcounter{magicrownumbers}
\newcommand\rownumber{\stepcounter{magicrownumbers}\arabic{magicrownumbers}}

\section {Implementation}
\label{sec:impl}




Let $P=$\prog be the input program.  First, we transform $P$ to
$\bp{P}$ by adding new \emph{counter} variables for the loops of $P$
and adding extra constraints to the transitions to count the number of
iterations. Specifically, for each location $\ell$ we introduce a counter
variable $c_\ell$ and a bounding variable $b_\ell$. Let $C$ and $B$ be
the sets of all counter and bounding variables, respectively, and
$\bd: C \to B$ be the bijection defined as $\bd(c_\ell) = b_\ell$. We
define $\bp{P} \equiv $ \quintuple {$L$} {$\ell^o$} {$\ell^e$} {$V
  \cup C \cup B$} {$\tau \land \tau_B$}, where $\tau_B(\ell_1, \ell_2)
= \bigand X(\ell_1, \ell_2)$ and $X(\ell_1, \ell_2)$ is the smallest
set satisfying the following conditions: (a) if $\ell_1\to \ell_2$ is
a back-edge, then $\left( 0 \le c_{\ell_2}' \;\land\; c_{\ell_2}' =
c_{\ell_2}-1 \;\land\; c_{\ell_2} \le b_{\ell_2} \right) \in X(\ell_1,
\ell_2)$, (b) else, if $\ell_1 \to \ell_2$ exits the loop headed by
$\ell_k$, then $\left( c_{\ell_k}=0 \right) \in X(\ell_1, \ell_2)$ and (c)
otherwise, if $\ell_1 \to \ell_2$ is a transition inside the loop headed by
$\ell_k$, then $\left( c'_{\ell_k} = c_{\ell_k} \right) \in X(\ell_1, \ell_2)$.
%
In practice, we use optimizations to reduce the number of variables
and constraints\footnote{More details are in the appendix.}.

This transformation preserves safety as shown below (proof in Appendix).

\begin{lemma}
\label{lem:bp_sound}
$P$ is safe iff $\bp{P}$ is safe, \ie if $\bp{\cex} = \langle \bar{\ell},
\bar{s} \rangle$ is a counterexample to $\bp{P}$, projecting $\bar{s}$
onto $V$ gives a counterexample to $P$; if $\bp{\pf}$ is a
proof of $\bp{P}$, then $\pf = \lambda \ell \cdot \{\forall B \ge 0,C \ge 0 \such
\varphi ~\mid~ \varphi \in \bp{\pf}(\ell)\}$ is a safety proof for $P$.
\end{lemma}



%

In the rest of this section, we define our abstractions and
under-approximations of $\bp{P}$ and describe our implementation of
the different routines in Fig.~\ref{fig:code}.

\subheading {Abstractions}
Recall that $\tau(\bp{P}) = \tau \land \tau_B$. W.l.o.g., assume that $\tau$ is
transformed to
$\exists \assum \such \left( \tau_\assum \land \bigand \assum \right)$ for a
finite set of fresh Boolean variables $\assum$ that only appear negatively in
$\tau_\assum$. We refer to $\assum$ as 
\emph{assumptions} following SAT terminology~\cite{minisat}. Dropping some
assumptions from $\bigand \assum$ results in an abstract transition relation,
\ie $\exists \assum \such \left( \tau_\assum \land
\bigand \hat{\assum} \right)$ is an abstraction of $\tau$ for $\hat{\assum}
\subseteq \assum$, denoted $\hat{\tau}(\hat{\assum})$. Note that
$\hat{\tau}(\hat{\assum}) = \tau_\assum
[\hat{\assum} \leftarrow \top, \assum \setminus \hat{\assum} \leftarrow \bot]$.
The only abstractions of $\bp{P}$ we consider are the ones which abstract $\tau$
and keep $\tau_B$ unchanged. That is, every abstraction $\hat{P}$ of $\bp{P}$ is
such that $\bp{P} \under_\id \hat{P}$ with $\tau(\hat{P}) = \hat{\tau}(\hat{\assum}) \land
\tau_B$ for some $\hat{\assum} \subseteq \assum$. Moreover, a subset
$\hat{\assum}$ of $\assum$  induces an abstraction of $\bp{P}$,
denoted $\bp{P}(\hat{\assum})$.

\subheading {Under-approximations} An under-approximation is induced
by a subset of assumptions $\hat{\assum} \subseteq \assum$, which
identifies the abstraction $\bp{P}(\hat{\assum})$, and a mapping
$\bvals : B \to \nat$ from $B$ to natural numbers, which bounds the
number of iterations of every loop in $\bp{P}$.  The
under-approximation, denoted $U(\hat{\assum},\bvals)$, satisfies
$U(\hat{\assum},\bvals) \sunder_\id \bp{P}(\hat{\assum})$, with
$\tau(U(\hat{\assum},\bvals)) = \hat{\tau}(\hat{\assum}) \land
\tau_B(\bvals)$ where $\tau_B(\bvals)$ is obtained from $\tau_B$ by
strengthening all transitions with $\bigand_{b \in B} b \le
\bvals(b)$.

\subheading {\textsc{Solve}}
We implement \solver (see Fig.~\ref{fig:code}) by transforming the
decidable under-approximation $U$, after restricting by the invariants to
$U_\invar$, to Horn-SMT~\cite{gpdr} (the input
format of $\mu Z$) and passing the result to \muz. Note that this
intentionally limits the power of \muz to solve only decidable problems.
In Section~\ref{sec:results}, we compare \algo \mbox{with~unrestricted~\muz}.

%

\begin{figure}[t]
\centering
\scriptsize
\begin{tabular}{l|ll@{\makebox[3em][r]{(\rownumber)\space}}}

\multirow{6}{*}{\emph{Global}}
& \quad \multirow{2}{*}{\emph{Trans}}
& \quad $\rtag {i} {j} \implies \tau_\assum (\ell_i, \ell_j) \land \tau_B (\ell_i, \ell_j)$,
	\quad $\ell_i, \ell_j \in L$ \\

& & \quad $\ltag {i} \implies \bigor_j \rtag {j} {i}$,
	\quad $\ell_i \in L$ \\
\\
& \quad \multirow{2}{*}{\emph{Invars}}
& \quad $\left( \bigor_j \rtag{i}{j} \right) \implies \varphi$,
	\quad $\ell_i \in L$, $\varphi \in \invar (\ell_i)$ \\
& & \quad $\ltag{i} \implies \varphi'$,
	\quad $\ell_i \in L$, $\varphi \in \invar (\ell_i)$ \\

\\
\hline
\\

\multirow{9}{*}{\emph{Local}}
& \quad \multirow{2}{*}{\emph{Lemmas}}
& \quad $\bigand_{\ell_i \in L, \varphi \in \pi(\ell_i)}
		\left( \mathcal{A}_{\ell_i, \varphi} \implies
		\left( \left( \bigor_j \rtag{i}{j} \right) \implies \varphi \right) \right)$ \\
& & \quad $\neg \bigand_{\ell_i \in L, \varphi \in \pi(\ell_i)}
		\left( \mathcal{B}_{\ell_i, \varphi} \implies
		\left( \ltag{i} \implies \varphi' \right) \right)$ \\
\\
& \quad \multirow{2}{*}{\emph{Assump. Lits}}
& \quad $\mathcal{A}_{\ell, \varphi}$, \quad $\ell \in L$, $\varphi \in \pi(\ell)$ \\
& & \quad $\neg \mathcal{B}_{\ell, \varphi}$, \quad $\ell \in L$, $\varphi \in \pi(\ell)$ \\
\\
& \quad \emph{Concrete}
& \quad $\Sigma$ \\
\\
& \quad \emph{Bound Vals}
& \quad $b \le \bvals(b)$, \quad $b \in B$ \\
\end{tabular}
\caption{Constraints used in our implementation of \algo.}
\label{fig:cons}
\end{figure}

We implement the routines of \algo in Fig.~\ref{fig:code} by
maintaining a set of constraints $\mathcal{C}$ as shown in
Fig.~\ref{fig:cons}. Initially, $\mathcal{C}$ is {\em Global}.
\emph{Trans} encodes the transition relation of $\bp{P}$, using fresh
Boolean variables for transitions and locations ($\rtag{i}{j}$,
$\ltag{i}$, respectively) enforcing that a location is reachable only
via one of its (incoming) edges. Choosing an abstract or concrete
transition relation is done by adding a subset of $\assum$ as additional
constraints.  \emph{Invars} encodes currently known
invariants.  They approximate the reachable states by adding constraints for
every invariant at a location in terms of current-state variables (3)
and next-state variables (4).  The antecedent in (3) specifies that at
least one transition from $\ell_i$ has been taken implying that the
current location is $\ell_i$ and the antecedent in (4) specifies that
the next location is $\ell_i$.

$\mathcal{C}$ is modified by each routine as needed by adding and
retracting some of the \emph{Local} constraints (see
Fig.~\ref{fig:cons}) as discussed below.

For a set of \emph{assumption literals}
$\mathcal {A}$,
let \textsc{Sat}$(\mathcal{C}, \mathcal{A})$ be a function that checks whether
$\mathcal{C} \cup \mathcal{A}$
is satisfiable, and if not, returns an \emph{unsat
core} $\hat{\mathcal{A}} \subseteq \mathcal{A}$ such that $\mathcal{C} \cup
\hat{\mathcal{A}}$ is unsatisfiable. 

In the rest of the section, we assume that $\pf$ is a safety proof of
$U_\invar(\hat{\assum},\bvals)$.

\subheading {\textsc{InitU}} The initial under-approximation is
$U(\assum, \lambda b\in B \such 0)$. 

\noindent
\textbf {\textsc{ExtractInvs}} is implemented by
\textsc{ExtractInvsImpl} shown in Fig.~\ref{fig:impl_code}. It extracts
a \emph{Maximal Inductive Subset} (MIS) of the lemmas in $\pf$
w.r.t. the concrete transition relation $\tau \land \tau_B$ of
$\bp{P}$.  First, the constraints \emph{Concrete} in
Fig.~\ref{fig:cons} are added to~$\mathcal{C}$, including all of
$\assum$. Second, the constraints \emph{Lemmas} in Fig.~\ref{fig:cons}
are added to~$\mathcal{C}$, where fresh Boolean variables
$\mathcal{A}_{\ell,\varphi}$ and $\mathcal{B}_{\ell,\varphi}$ are used to mark every lemma
$\varphi$ at every location $\ell \in L$. This encodes the negation of
the inductiveness condition of a safety proof (see
Def.~\ref{def:proof}).

The MIS of $\pf$ corresponds to the \emph{maximal} subset $I \subseteq
\{\mathcal{A}_{\ell,\varphi}\}_{\ell,\varphi}$ such that $\mathcal{C} \cup I \cup \{\neg
\mathcal{B}_{\ell,\varphi} \mid \mathcal{A}_{\ell,\varphi} \not \in I\}$ is unsatisfiable.
$I$ is computed by \textsc{ExtractInvsImpl} in
Fig.~\ref{fig:impl_code}. Each iteration of \textsc{ExtractInvsImpl}
computes a \emph{Minimal Unsatisfiable Subset} (MUS) to identify (a minimal set
of) more non-inductive lemmas (lines~3--6). $M$, on line~4, indicates the
cumulative set of non-inductive lemmas and $X$, on line~5, indicates all the other
lemmas.  $\textsc{Mus}(\mathcal{C},T,V)$ in Fig.~\ref{fig:impl_code}
iteratively computes a minimal subset, $R$, of $V$ such that $\mathcal{C}
\cup T \cup R$ is unsatisfiable.


\begin{figure}[t]
	\begin{subfigure}[t]{.57\textwidth}
	\begin{algorithm}[H]
	\scriptsize
	\DontPrintSemicolon

	{\sc ExtractInvsImpl}$(\mathcal{C}, \{\mathcal{A}_{\ell, \varphi}\}_{\ell,\varphi},
\{\mathcal{B}_{\ell,\varphi}\}_{\ell,\varphi})$\;
	\Begin {
\nl		$M$ := $\emptyset$, $X$ := $\{\mathcal{A}_{\ell,\varphi}\}_{\ell,\varphi}$, $Y$ :=
$\{\neg \mathcal{B}_{\ell,\varphi}\}_{\ell,\varphi}$\;
\nl		$T$ := $X$\;
\nl		\While {$(S := \textsc{Mus}(\mathcal{C},T,Y)) \neq \emptyset $ } {
\nl			$M$ := $M \cup S$, $Y$ := $Y \setminus M$\;
\nl			$X$ := $\{\mathcal{A}_{\ell,\varphi} ~|~ \neg
\mathcal{B}_{\ell,\varphi} \in Y\}$\;
\nl			$T$ := $X \cup M$\;
		}
\nl		\Return $X$
	}
	\end{algorithm}
	\end{subfigure}
	\begin{subfigure}[t]{.45\textwidth}
	\begin{algorithm}[H]
	\scriptsize
	\DontPrintSemicolon
    	\setcounter{AlgoLine}{7}

	{\sc Mus}$(\mathcal{C}, T, V)$\;
	\Begin {
\nl		$R$ := $\emptyset$\;
\nl		\While {{\sc Sat}$(\mathcal{C}, T \cup R)$} {
\nl			$m$ := {\sc GetModel}$(\mathcal{C}, T \cup R)$\;
\nl			$R$ := $R \cup \{v \in V ~|~ m(\neg v)\}$\;
		}
\nl		\Return $R$\;
	}
	\end{algorithm}
	\end{subfigure}
\caption {Our implementation of \textsc{ExtractInvs} of Fig.~\ref{fig:code}.}
\label {fig:impl_code}
\end{figure}

\noindent
\textbf{\textsc{Abstract}} finds a $\hat{\assum}_1 \subseteq \assum$
such that $U_\invar(\hat{\assum}_1,\bvals)$ is safe with proof $\pf$. The
constraints \emph{Lemmas} in Fig.~\ref{fig:cons} are added to $\mathcal{C}$
to encode the negation of the conditions in
Definition~\ref{def:proof}. Then, the constraints in \emph{Bound Vals}
in Fig.~\ref{fig:cons} are added to $\mathcal{C}$ to encode the
under-approximation. This reduces the check for $\pf$ to be a safety
proof to that of unsatisfiability of a formula. Finally, {\sc
  Sat}$(\mathcal{C}, \assum \cup \{\mathcal{A}_{\ell,\varphi}\}_{\ell,\varphi} \cup
\{\mathcal{B}_{\ell,\varphi}\}_{\ell,\varphi})$ is invoked. As $\mathcal{C}$ is
unsatisfiable assuming $\assum$ and using all the lemmas (since $\pf$
proves $U_\invar(\hat{\assum}, \bvals)$), it returns an unsat core.
Projecting the core onto $\assum$ gives us $\hat{\assum}_1 \subseteq
\assum$ which identifies the new abstraction and, together with
$\bvals$, the corresponding new under-approximation. The minimality of
$\hat{\assum}_1$ depends on the algorithm for extracting an unsat
core, which is part of the SMT engine of Z3 in our case. In practice,
we use a \emph{Minimal Unsatisfiable Subset} (MUS) algorithm to find a
minimal $\hat{\assum}_1$. As we treat
$\{\mathcal{A}_{\ell,\varphi}\}_{\ell,\varphi}$ and
$\{\mathcal{B}_{\ell,\varphi}\}_{\ell,\varphi}$ as assumption literals, this
also corresponds to using only the necessary lemmas during
abstraction.

\subheading{\textsc{NextU}}
Given the current valuation $\bvals$ and the new abstraction $\hat{\assum}$,
this routine returns $U(\hat{\assum}, \lambda b \in B  \such \bvals(b) + 1)$.


\subheading {\textsc{Refine} and \textsc{IsFeasible}} Let
$U_\invar(\hat{\assum},\bvals)$ be unsafe with a counterexample $\cex$.  We
create a new set of constraints $\mathcal{C}_\cex$ corresponding to the
unrolling of $\tau_\assum \land \tau_B$ along the control path of
$\cex$ and check {\sc Sat}$(\mathcal{C}_\cex, \assum)$.  If the path is
feasible in $\bp{P}$, we find a counterexample to safety in
$\bp{P}$. Otherwise, we obtain an unsat core $\hat{\assum}_1 \subseteq \assum$ and
refine the abstraction to $\hat{\assum} \cup \hat{\assum}_1$. The
under-approximation is refined accordingly with the same $\bvals$.

We conclude the section with a discussion of the implementation
choices.  \textsc{NextU} is implemented by incrementing all bounding
variables uniformly. An alternative is to increment the bounds only for the loops
whose invariants are not inductive (\eg~\cite{ufo,impact}).
However, we leave the exploration of such
strategies for future. Our use of \muz is sub-optimal since each
call to \solver requires constructing a new Horn-SMT problem. This
incurs an unnecessary pre-processing overhead that can be eliminated
by a tighter integration with \muz. For \textsc{Abstract} and
\textsc{ExtractInvs}, we use a single SMT-context with a single copy
of the transition relation of the program (without unrolling it). The
context is preserved across iterations of \algo. Constraints specific
to an iteration are added and retracted using the incremental solving
API of Z3. This is vital for performance. For \textsc{Refine} and
\textsc{IsFeasible}, we unroll the transition relation of the program
along the control path of the counterexample trace returned by \muz. We experimented with
an alternative implementation that instead validates each individual
step of the counterexample using the same global context as
\textsc{Abstract}. While this made each refinement step faster, it
increased the number of refinements, becoming inefficient overall.









\newcommand{\tout}{TO}
\newcommand{\mout}{MO}

\section{Experiments}
\label{sec:results}

We implemented \algo in Python using Z3~v4.3.1 (with a few
modifications to Z3 API\footnote{Our changes are being incorporated
  into Z3, and will be available in future versions.}). The
implementation and complete experimental results are available
at~\url{http://www.cs.cmu.edu/~akomurav/projects/spacer/home.html}.


\paragraph{Benchmarks.} We evaluated \algo on the benchmarks from the
\emph{systemc}, \emph{product-lines}, \emph{device-drivers-64} and
\emph{control-flow-integers} categories of SV-COMP'13. Other
categories require bit-vector and heap reasoning that are not
supported by \algo. We used the front-end of
\textsc{UFO}~\cite{ufo_svcomp} to convert the benchmarks from C to the
Horn-SMT format of $\mu Z$.


Overall, there are 1,990 benchmarks (1,591 SAFE, and 399
UNSAFE); 1,382 are decided by the \textsc{UFO} front-end that uses
common compiler optimizations to reduce the problem. This left 608
benchmarks (231 SAFE, and 377 UNSAFE).

\begin{figure}[p]
    \centering
    \includegraphics[scale=.8]{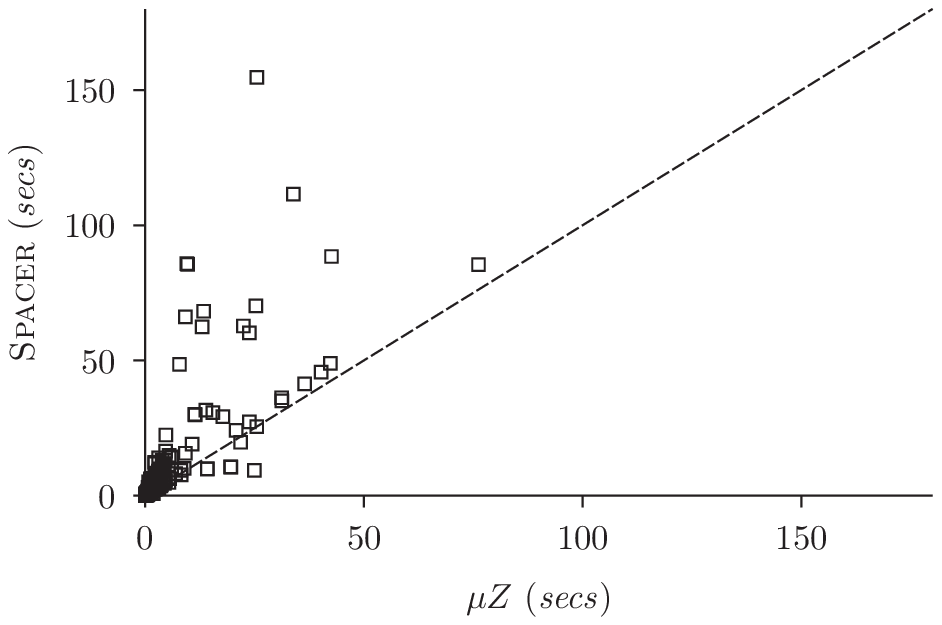}
    \caption{\algo vs. \muz for UNSAFE benchmarks.}
    \label{fig:down_sv_pba_unsafe}
    \vspace{0.4in}
    \centering
    \includegraphics[scale=.8]{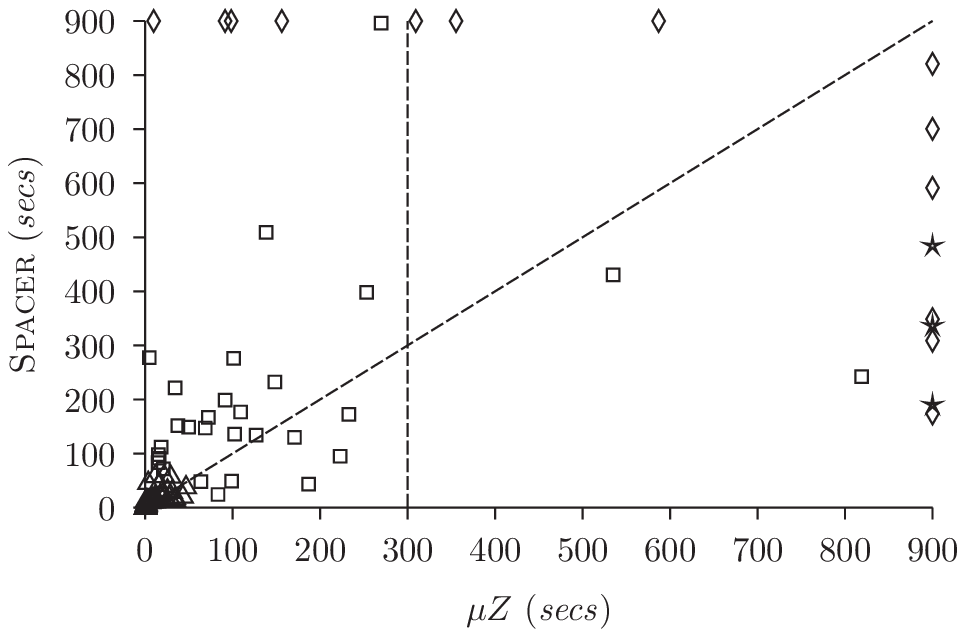}
    \caption{\algo vs. \muz for SAFE benchmarks.}
    \label{fig:down_safe}
    \vspace{0.4in}
    \centering
    \includegraphics[scale=.8]{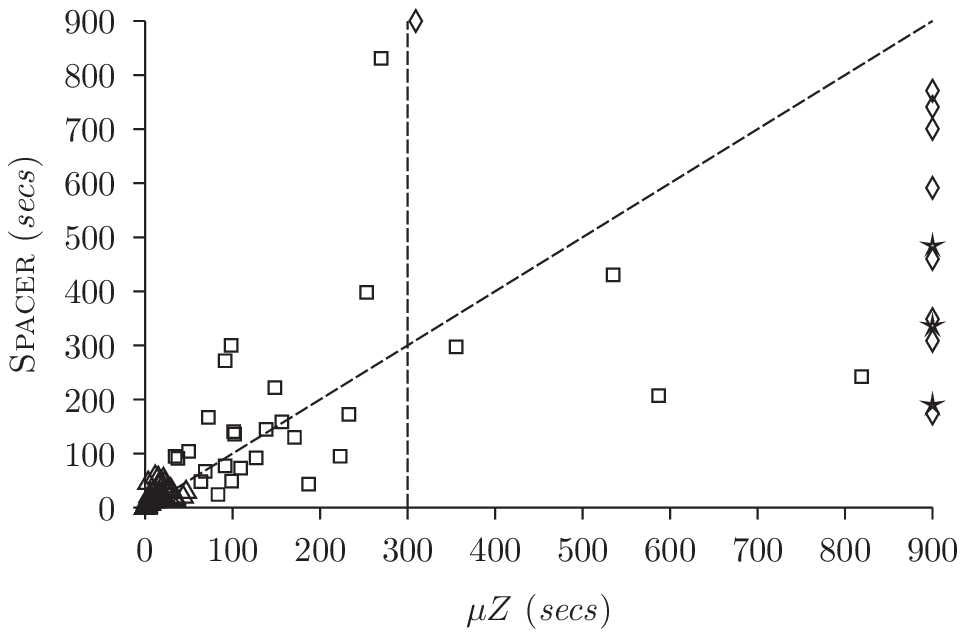}
    \caption{Best of the three variants of \algo vs. \muz for SAFE benchmarks.}
    \label{fig:down_sv_pba_safe}
\end{figure}

For the UNSAFE benchmarks, 369 cases are solved by both \muz and
\algo; in the remaining 8 benchmarks, 6 are solved by neither tool,
and 2 are solved by \muz but not by \algo. Fig.~\ref{fig:down_sv_pba_unsafe} shows a
scatter plot comparing \algo with \muz on these 369 cases. Note that, even
though abstraction did not help for these benchmarks, hurting
significantly in some cases, the benchmarks are easy with \algo needing at
most $3$ minutes each.

For the SAFE benchmarks, see Fig.~\ref{fig:down_safe} for a scatter plot
comparing \algo with \muz. 176 cases are solved in under a minute by both
tools (see the dense set of triangles in the lower left corner of the figure). For them, the
difference between \algo and $\mu Z$ is not significant to be meaningful.
Of the remaining 55 hard benchmarks 42 are solved by either \muz, \algo or both with a
time limit of 15 minutes and 2GB of memory. The rest remain unsolved. All
experiments were done on an Intel\textregistered~Core$^\text{TM}$2 Quad
CPU of 2.83GHz and 4GB of RAM.

\paragraph{Results.} Table~\ref{tab:results} shows the experimental
results on the 42 solved benchmarks needing more than a minute of running time.
The~$t$ columns under \muz and
\algo show the running times in seconds with `TO' indicating a
time-out and a `MO' indicating a mem-out. The best times are
highlighted in bold. The corresponding scatter plot in
Fig.~\ref{fig:down_safe} shows that the results are mixed for a time bound
of 300 seconds (5 minutes). But beyond 5 minutes, abstraction really helps with many benchmarks
solved by \algo when \muz runs out of time (time-outs are indicated by diamonds).
The couple of benchmarks where \algo runs out of time become better than \muz
using a different setting, as discussed later.
Overall, abstraction helps for \emph{hard}
benchmarks. Furthermore, in {\tt elev\_13\_22}, {\tt elev\_13\_29} and {\tt elev\_13\_30},
\algo is successful even though $\mu Z$ runs out of memory, showing a
clear advantage of abstraction (this corresponds to the stars in the far right
of Fig.~\ref{fig:down_safe}). Note that \texttt{gcnr}, under
\emph{misc}, in the table is the example from Fig.~\ref{fig:abs_helps}.


\begin{table}[t]
\centering
\scriptsize
\begin{tabular}{p{.5\textwidth}p{.5\textwidth}}
\begin{tabular}{|l|r||r|r|r|r||r|r|}
\hline
\multicolumn{1}{|c|}{\emph{Benchmark}} & \multicolumn{1}{c||}{$\mu Z$} & \multicolumn{6}{c|}{\algo} \\
\cline {2-8}
& $t$ & $t$ & $B$ & $a_f$ & $a_m$ & $t_p$ & $B_p$\\
& ({\em sec}) & ({\em sec}) & & $(\%)$ & $(\%)$ & ({\em sec}) & \\
\hline

\multicolumn{8}{|c|}{\em systemc} \\
\hline
{\tt pipeline} & $224$ 
               & ${\bf 120}$ & $4$ & $33$ & $33$ & $249$ & $4$ \\
{\tt tk\_ring\_06} & $64$ 
                   & ${\bf 48}$ & $2$ & $59$ & $59$ & $65$ & $2$ \\
{\tt tk\_ring\_07} & $69$ 
                   & $120$ & $2$ & $59$ & $59$ & $\dagger \bf 67$ & $2$ \\
{\tt tk\_ring\_08} & $232$ 
                   & ${\bf 158}$ & $2$ & $57$ & $57$ & $358$ & $2$ \\
{\tt tk\_ring\_09} & $817$ 
                   & ${\bf 241}$ & $2$ & $59$ & $59$ & $266$ & $2$ \\
{\tt mem\_slave\_1} & $536$ 
                    & ${\bf 430}$ & $3$ & $24$ & $34$ & $483$ & $2$ \\
{\tt toy} & \tout 
          & $822$ & $4$ & $32$ & $44$ & $\dagger \bf 460$ & $4$ \\

{\tt pc\_sfifo\_2} & $\bf 73$ 
                   & $137$ & $2$ & $41$ & $41$ & \tout & $-$ \\
\hline

\multicolumn{8}{|c|}{\em product-lines} \\
\hline
{\tt elev\_13\_21} & \tout 
                   & ${\bf 174}$ & $2$ & $7$ & $7$ & \tout & $-$ \\
{\tt elev\_13\_22} & \mout 
                   & ${\bf 336}$ & $2$ & $9$ & $9$ & $624$ & $4$ \\
{\tt elev\_13\_23} & \tout 
                   & ${\bf 309}$ & $4$ & $6$ & $14$ & \tout & $-$ \\
{\tt elev\_13\_24} & \tout 
                   & ${\bf 591}$ & $4$ & $9$ & $9$ & \tout & $-$ \\
{\tt elev\_13\_29} & \mout 
                   & ${\bf 190}$ & $2$ & $6$ & $10$ & \tout & $-$ \\

{\tt elev\_13\_30} & \mout 
                   & ${\bf 484}$ & $3$ & $11$ & $13$ & \tout & $-$ \\

{\tt elev\_13\_31} & \tout 
                   & ${\bf 349}$ & $4$ & $8$ & $17$ & \tout & $-$ \\
{\tt elev\_13\_32} & \tout 
                   & ${\bf 700}$ & $4$ & $9$ & $9$ & \tout & $-$ \\

{\tt elev\_1\_21} & $\bf 102$ 
                  & $136$ & $11$ & $61$ & $61$ & $161$ & $11$ \\
{\tt elev\_1\_23} & $\bf 101$ 
                  & $276$ & $11$ & $61$ & $61$ & $\dagger 140$ & $11$ \\
{\tt elev\_1\_29} & $92$ 
                  & $199$ & $11$ & $61$ & $62$ & $\dagger \bf 77$ & $11$ \\
{\tt elev\_1\_31} & $127$ 
                  & $135$ & $11$ & $62$ & $62$ & $\dagger \bf 92$ & $11$ \\

{\tt elev\_2\_29} & $\bf 18$ 
                  & $112$ & $11$ & $56$ & $56$ & $\dagger 26$ & $11$ \\
{\tt elev\_2\_31} & $\bf 16$ 
                  & $91$ & $11$ & $57$ & $57$ & $\dagger 22$ & $11$ \\
\hline
\end{tabular} &
\begin{tabular}{|l|r||r|r|r|r||r|r|}
\hline
\multicolumn{1}{|c|}{\emph{Benchmark}} & \multicolumn{1}{c||}{$\mu Z$} & \multicolumn{6}{c|}{\algo} \\
\cline {2-8}
& $t$ & $t$ & $B$ & $a_f$ & $a_m$ & $t_p$ & $B_p$\\
& ({\em sec}) & ({\em sec}) & & $(\%)$ & $(\%)$ & ({\em sec}) & \\
\hline

\multicolumn{8}{|c|}{\em ssh} \\
\hline
{\tt s3\_clnt\_3} & $109$ 
                  & $*90$ & $12$ & $13$ & $13$ & $\bf 73$ & $12$ \\
{\tt s3\_srvr\_1} & $187$ 
                  & ${\bf 43}$ & $9$ & $18$ & $18$ & $661$ & $25$ \\
{\tt s3\_srvr\_2} & $587$ 
                  & $*{\bf 207}$ & $14$ & $3$ & $7$ & $446$ & $15$ \\
{\tt s3\_srvr\_8} & $99$ 
                  & ${\bf 49}$ & $13$ & $18$ & $18$ & \tout & $-$ \\
{\tt s3\_srvr\_10} & $83$ 
                   & ${\bf 24}$ & $9$ & $17$ & $17$ & $412$ & $21$ \\
{\tt s3\_srvr\_13} & $355$ 
                   & $*{\bf 298}$ & $15$ & $8$ & $8$ & $461$ & $15$ \\

{\tt s3\_clnt\_2} & $\bf 34$
                  & $*124$ & $13$ & $13$ & $13$ & $\dagger 95$ & $13$ \\
{\tt s3\_srvr\_12} & $\bf 21$
                  & $*64$ & $13$ & $8$ & $8$ & $54$ & $13$ \\
{\tt s3\_srvr\_14} & $\bf 37$
                  & $*141$ & $17$ & $8$ & $8$ & $\dagger 91$ & $17$ \\

{\tt s3\_srvr\_6} & $\bf 98$ 
                   & \tout & $-$ & $-$ & $-$ & $\dagger 300$ & $25$ \\
{\tt s3\_srvr\_11} & $\bf 270$ 
                   & $896$ & $15$ & $14$ & $18$ & $831$ & $13$ \\
{\tt s3\_srvr\_15} & $\bf 309$ 
                   & \tout & $-$ & $-$ & $-$ & \tout & $-$ \\
{\tt s3\_srvr\_16} & $\bf 156$ 
                   & $*263$ & $21$ & $8$ & $8$ & $\dagger 159$ & $21$ \\
\hline

\multicolumn{8}{|c|}{\em ssh-simplified} \\
\hline
{\tt s3\_srvr\_3} & $171$ 
                  & $130$ & $11$ & $21$ & $21$ & $\bf 116$ & $12$ \\

{\tt s3\_clnt\_3} & $\bf 50$
                  & $*139$ & $12$ & $17$ & $22$ & $\dagger 104$ & $13$ \\
{\tt s3\_clnt\_4} & $\bf 15$
                  & $*76$ & $12$ & $22$ & $22$ & $56$ & $13$ \\

{\tt s3\_clnt\_2} & $\bf 138$ 
                  & $509$ & $13$ & $26$ & $26$ & $\dagger 145$ & $13$ \\
{\tt s3\_srvr\_2} & $\bf 148$ 
                  & $232$ & $12$ & $16$ & $23$ & $222$ & $15$ \\
{\tt s3\_srvr\_6} & $\bf 91$ 
                  & \tout & $-$ & $-$ & $-$ & $\dagger 272$ & $25$ \\
{\tt s3\_srvr\_7} & $\bf 253$ 
                  & $398$ & $10$ & $20$ & $26$ & $764$ & $10$ \\
\hline

\multicolumn{8}{|c|}{\em misc} \\
\hline
{\tt gcnr} & \tout
           & $56$ & $26$ & $81$ & $95$ & $\bf 50$ & $25$ \\
\hline
\end{tabular}
\end{tabular}
\vspace{0.1in}
\caption{Comparison of \muz and \algo. $t$ and $t_p$ are running times in
seconds; $B$  and $B_p$ are the final values of the bounding variables; $a_f$
and $a_m$ are the fractions
  of assumption variables in the final and maximal abstractions, respectively. }
\label{tab:results}
\end{table}

The $B$ column in the table shows the final values of the loop bounding
variables under the mapping \bvals, \ie the maximum number of loop iterations (of any loop) that was necessary for the
final safety proof. Surprisingly, they are very small in many of the
hard instances in \emph{systemc} and \emph{product-lines} categories.

Columns $a_f$ and $a_m$ show the sizes of the final and maximal
abstractions, respectively, measured in terms of the number of the
original constraints used. Note that this only corresponds to the
\emph{syntactic}  abstraction (see
Section~\ref{sec:details}). The final abstraction done by \algo is
very aggressive. Many constraints are irrelevant with often, more than
50\% of the original constraints abstracted away.  Note that this
is in addition to the aggressive property-independent abstraction done by
the \textsc{UFO} front-end. Finally, the difference between $a_f$ and
$a_m$ is insignificant in all of the benchmarks.


Another approach to \textsc{Abstract} is to restrict abstraction to
state-variables by making assignments to some next-state variables
non-deterministic, as done by Vizel et al.~\cite{orna_fmcad12} in a similar
context. This was especially
effective for \emph{ssh} and \emph{ssh-simplified} categories -- see
the entries marked with `*' under column $t$.

An alternative implementation of \textsc{Refine} is to concretize the
under-approximation (by refining $\hat{\Sigma}$ to $\Sigma$) whenever
a spurious counterexample is found. This is analogous to Proof-Based
Abstraction (PBA)~\cite{pba1} in hardware verification. Run-time for PBA and
the corresponding final values of the bounding variables are shown in columns $t_p$ and $B_p$ of
Table~\ref{tab:results}, respectively. While this results in more time-outs,
it is significantly better in 14 cases (see the entries marked with `$\dagger$'
under column $t_p$), with 6 of them comparable to \muz and 2 (\viz \texttt{toy} and
\mbox{\texttt{elev\_1\_31}}) significantly better than~\muz.

See Fig.~\ref{fig:down_sv_pba_safe} for a scatter plot using the best running
times for \algo of all the three variants described above.






We conclude this section by comparing our results with
\textsc{UFO}~\cite{ufo_svcomp} --- the winner of the 4 categories at
SV-COMP'13. The competition version of \textsc{UFO} runs several
engines in parallel, including engines based on Abstract
Interpretation, Predicate Abstraction and \ubmc with
interpolation. \textsc{UFO} outperforms \algo and \muz in \emph{ssh}
and \emph{product-lines} categories by an order of magnitude. They are
difficult for \ubmc, but easy for Abstract Interpretation and
Predicate Abstraction, respectively. Even so, note that \algo finds really small
abstractions for these categories upon termination. However, in the \emph{systemc}
category both \algo and \muz perform better than \textsc{UFO} by
solving hard instances (\eg \texttt{tk\_ring\_08} and
\texttt{tk\_ring\_09}) that are not solved by any tool in the
competition. Moreover, \algo is faster than \muz.  Thus, while \algo
itself is not the best tool for all benchmarks, it is a valuable
addition to the state-of-the-art verification engines.






\section{Related work}

There is a large body of work on \ubmc approaches both in hardware and
software verification. In this section, we briefly survey the most
related work.

The two most prominent approaches to \ubmc combine BMC with
interpolation (e.g.,~\cite{ufo,mcmillan03,impact}) or with inductive
generalization (e.g.,~\cite{ic3,smc_ic3,pdr,gpdr}). Although our implementation
of \algo is based on inductive generalization (the engine of \muz),
it can be implemented on top of an interpolation-based engine as well.

Proof-based Abstraction (PBA) was first introduced in hardware
verification to leverage the power of SAT-solvers to focus on relevant
facts~\cite{pba2,pba1}. Over the years, it has been combined with
CEGAR~\cite{pba+cba,ar+sat}, interpolation~\cite{ar+sat,ar+interp}, and
PDR~\cite{jason_fmcad12}. To the best of our knowledge, \algo is the first
application of PBA to software verification.

The work of Vizel et al.~\cite{orna_fmcad12}, in hardware
verification, that extends PDR with abstraction is closest to
ours.
However, \algo is not tightly coupled with PDR, which
makes it more general, but possibly, less efficient. Nonetheless,
\algo allows for a rich space of abstractions,
whereas Vizel et al. limit themselves to state variable abstraction.

\newcommand{\ufo}{\textsc{Ufo}\xspace}

Finally, \ufo~\cite{ufo,vinta} also combines abstraction with \ubmc, but in an
orthogonal way. In \ufo, abstraction is used to guess the depth of
unrolling (plus useful invariants), BMC to detect
counterexamples, and interpolation to synthesize safe inductive
invariants. While \ufo performs well on many competition benchmarks,
combining it with \algo will benefit on the hard ones.



\newcommand{\vinta}{\textsc{Vinta}\xspace}

\section{Conclusion}
\label{sec:conclusion}
In this paper, we present an algorithm, \algo, that combines
Proof-Based Abstraction (PBA) with CounterExample Guided Abstraction
Refinement (CEGAR) for verifying safety properties of sequential
programs. To our knowledge, this is the first application of PBA to
software verification.  Our abstraction technique combines localization with
invariants about the program. It is interesting to explore
alternatives for such a \emph{semantic} abstraction.

While our presentation is restricted to non-recursive sequential
programs, the technique can be adapted to solving the more general
Horn Clause Satisfiability problem and extended to
verifying recursive and concurrent programs~\cite{hsfc}.

We have implemented \algo in Python using Z3 and its GPDR engine
\muz. The current implementation is an early prototype. It is not
heavily optimized and is not tightly integrated with
\muz. Nonetheless, the experimental results on 4 categories of the 2nd
Software Verification Competition show that \algo improves on both
\muz and the state-of-the-art.


\subsubsection{Acknowledgment.}  We thank Nikolaj Bj{\o}rner for many
helpful discussions and help with \muz and the anonymous reviewers for
insightful comments.



\begin{thebibliography}{10}

\bibitem{vinta}
A.~Albarghouthi, A.~Gurfinkel, and M.~Chechik.
\newblock {Craig Interpretation}.
\newblock In {\em SAS}, 2012.

\bibitem{ufo}
A.~Albarghouthi, A.~Gurfinkel, and M.~Chechik.
\newblock {From Under-Approximations to Over-Approximations and Back}.
\newblock In {\em TACAS}, 2012.

\bibitem{ufo_svcomp}
A.~Albarghouthi, A.~Gurfinkel, Y.~Li, S.~Chaki, and M.~Chechik.
\newblock {UFO: Verification with Interpolants and Abstract Interpretation -
  (Competition Contribution)}.
\newblock In {\em TACAS}, 2013.

\bibitem{pba+cba}
N.~Amla and K.~L. McMillan.
\newblock {A Hybrid of Counterexample-Based and Proof-Based Abstraction}.
\newblock In {\em FMCAD}, pages 260--274, 2004.

\bibitem{ar+sat}
N.~Amla and K.~L. McMillan.
\newblock {Combining Abstraction Refinement and SAT-Based Model Checking}.
\newblock In {\em TACAS}, 2007.

\bibitem{babic07}
D.~Babic and A.~J. Hu.
\newblock {Structural Abstraction of Software Verification Conditions}.
\newblock In {\em CAV}, 2007.

\bibitem{slam}
T.~Ball, R.~Majumdar, T.~Millstein, and S.~K. Rajamani.
\newblock {Automatic Predicate Abstraction of C Programs}.
\newblock {\em SIGPLAN Not.}, 36(5):203--213, 2001.

\bibitem{bmc}
A.~Biere, A.~Cimatti, E.~M. Clarke, O.~Strichman, and Y.~Zhu.
\newblock {Bounded Model Checking}.
\newblock {\em Advances in Computers}, 58:117--148, 2003.

\bibitem{wto}
F.~Bourdoncle.
\newblock {Efficient Chaotic Iteration Strategies with Widenings}.
\newblock In {\em Formal Methods in Programming and their Applications}, pages
  128--141, 1993.

\bibitem{ic3}
A.~R. Bradley.
\newblock {SAT-Based Model Checking without Unrolling}.
\newblock In {\em VMCAI}, 2011.

\bibitem{smc_ic3}
A.~Cimatti and A.~Griggio.
\newblock {Software Model Checking via IC3}.
\newblock In {\em CAV}, 2012.

\bibitem{cegar}
E.~M. Clarke, O.~Grumberg, S.~Jha, Y.~Lu, and H.~Veith.
\newblock {Counterexample-Guided Abstraction Refinement}.
\newblock In {\em CAV}, 2000.

\bibitem{z3}
L.~De~Moura and N.~Bj{\o}rner.
\newblock {Z3: An Efficient SMT Solver}.
\newblock In {\em TACAS}, 2008.

\bibitem{pdr}
N.~E{\'e}n, A.~Mishchenko, and R.~K. Brayton.
\newblock {Efficient Implementation of Property Directed Reachability}.
\newblock In {\em FMCAD}, pages 125--134, 2011.

\bibitem{minisat}
N.~E{\'e}n and N.~S{\"o}rensson.
\newblock {An Extensible SAT-solver}.
\newblock In {\em SAT}, 2003.

\bibitem{houdini}
C.~Flanagan and K.~R.~M. Leino.
\newblock {Houdini, an Annotation Assistant for ESC/Java}.
\newblock In {\em FME}, pages 500--517, 2001.

\bibitem{hsfc}
S.~Grebenshchikov, N.~P. Lopes, C.~Popeea, and A.~Rybalchenko.
\newblock {Synthesizing Software Verifiers from Proof Rules}.
\newblock In {\em PLDI}, pages 405--416, 2012.

\bibitem{mathsat}
A.~Griggio.
\newblock {A Practical Approach to Satisfiability Modulo Linear Integer
  Arithmetic}.
\newblock {\em JSAT}, 8:1--27, January 2012.

\bibitem{gulavani}
B.~S. Gulavani, S.~Chakraborty, A.~V. Nori, and S.~K. Rajamani.
\newblock {Automatically Refining Abstract Interpretations}.
\newblock In {\em TACAS}, 2008.

\bibitem{pba2}
A.~Gupta, M.~K. Ganai, Z.~Yang, and P.~Ashar.
\newblock {Iterative Abstraction using SAT-based BMC with Proof Analysis}.
\newblock In {\em ICCAD}, pages 416--423, 2003.

\bibitem{blast}
T.~A. Henzinger, R.~Jhala, R.~Majumdar, and G.~Sutre.
\newblock {Lazy Abstraction}.
\newblock {\em SIGPLAN Not.}, 37(1):58--70, 2002.

\bibitem{gpdr}
K.~Hoder and N.~Bj{\o}rner.
\newblock {Generalized Property Directed Reachability}.
\newblock In {\em SAT}, 2012.

\bibitem{jason_fmcad12}
A.~Ivrii, A.~Matsliah, H.~Mony, and J.~Baumgartner.
\newblock {IC3-Guided Abstraction}.
\newblock In {\em FMCAD}, 2012.

\bibitem{itp+pa}
R.~Jhala and K.~L. McMillan.
\newblock {A Practical and Complete Approach to Predicate Refinement}.
\newblock In {\em TACAS}, 2006.

\bibitem{ar+interp}
B.~Li and F.~Somenzi.
\newblock {Efficient Abstraction Refinement in Interpolation-Based Unbounded
  Model Checking}.
\newblock In {\em TACAS}, 2006.

\bibitem{mcmillan03}
K.~L. McMillan.
\newblock {Interpolation and SAT-Based Model Checking}.
\newblock In {\em CAV}, 2003.

\bibitem{impact}
K.~L. McMillan.
\newblock {Lazy Abstraction with Interpolants}.
\newblock In {\em CAV}, 2006.

\bibitem{pba1}
K.~L. McMillan and N.~Amla.
\newblock {Automatic Abstraction without Counterexamples}.
\newblock In {\em TACAS}, 2003.

\bibitem{duality}
K.~L. McMillan and A.~Rybalchenko.
\newblock {Solving Constrained Horn Clauses using Interpolation}.
\newblock Technical Report MSR-TR-2013-6, Microsoft Research, 2013.

\bibitem{orna_fmcad12}
Y.~Vizel, O.~Grumberg, and S.~Shoham.
\newblock {Lazy Abstraction and SAT-based Reachability for Hardware Model
  Checking}.
\newblock In {\em FMCAD}, 2012.

\end{thebibliography}

\newpage
\appendix

\section {Transformation of $P$ to $\bp{P}$}
The details of the transformation of an input program $P$ to $\bp{P}$ by
introducing counter variables is discussed below.

First, we construct a {\em Weak Topological Order} (WTO)~\cite{wto} of
$P$, which is a well-parenthesized total order of $L$, denoted $\wto$,
without two consecutive open brackets, satisfying the following
condition. Let the locations within a matching open-close bracket pair
constitute a {\em component} and let the smallest location w.r.t
$\wto$ in a component be its {\em head}. Let $\hds{\ell}$ be the
outside-in list of the heads of components containing $\ell$. Let
$\ell_1 \leq \ell_2 \equiv (\ell_1 = \ell_2 \lor \ell_1 <
\ell_2)$. Then,
\begin{equation} \label{eq:wto}
\forall \ell_i,\ell_j \in L \such \tau(\ell_i,\ell_j) \land
\ell_j \wwto \ell_i \implies \ell_j \in \hds{\ell_i}
\end{equation}


\begin{figure}[t]
\centering
\includegraphics[scale=1]{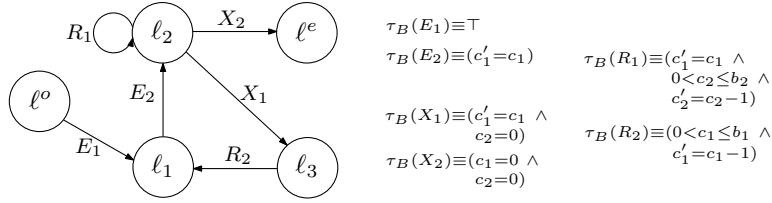}
\caption{Program with a nested loop and its corresponding {\em bounded}
transition constraints.}
\label{fig:nested_loops}
\end{figure}

Intuitively, $\wto$ is a total order of $L$ such that each component
identifies a loop in $P$, the head of a component identifies the entry
location of the loop and $\hds{\ell}$ denotes the outside-in list of
nested loops containing $\ell$. Condition~\eqref{eq:wto} says that a
{\em back-edge}, w.r.t $\wto$, leads to the head of a component
containing the source of the edge, denoting the start of a new
iteration of the corresponding loop. For example,
Fig.~\ref{fig:nested_loops} shows a program with two loops, an outer
loop \triple {$\ell_1$} {$\ell_2$} {$\ell_3$} and an inner loop
$\langle \ell_2 \rangle$. One possible WTO for this program is
``$\ell^o (\ell_1 (\ell_2) \ell_3) \ell^e$'' with $\ell_1$ and
$\ell_2$ as the heads of the two components. Without loss of
generality, assume that $\ell^o$ is always the smallest and $\ell^e$
is always the largest location of a WTO.

\subheading{Bound Variables}
Next, we introduce a set $C$ of rational variables, one per head of a component,
and the corresponding partial mapping $\ctr : L \partialto C$. Intuitively,
$\ctr(\ell)$ is the number of iterations (completed or remaining, depending on
whether we are counting up or down, respectively) of the component whose head
is $\ell$. Also, let $B$ be another set of rational variables, and
$\bd : C \to B$ be a bijection (i.e., $|B| = |C|$). Informally,
$\bd(c)$ denotes the upper bound of $c$. For example, in
Fig.~\ref{fig:nested_loops} we have $C = \{ c_1, c_2 \}$, $c_1 =
\ctr(\ell_1)$, $c_2 = \ctr(\ell_2)$, $B = \{ b_1, b_2 \}$, $\bd(c_1) =
b_1$, and $\bd(c_2) = b_2$. We construct a {\em bounded} program
$\bp{P} =$ \quintuple {$L$} {$\ell^o$} {$\ell^e$} {$V \cup C \cup B$}
{$\bp{\tau}$}, where $\forall \ell_i,\ell_j \in L \such
\bp{\tau}(\ell_i,\ell_j) = \tau(\ell_i,\ell_j) \land
\tau_B(\ell_i,\ell_j)$, and $\tau_B(\ell_i,\ell_j)$ is a set of
constraints defined as follows, assuming $c_j=\ctr(\ell_j)$ and
$b_j=\bd(c_j)$:


\emph{Entry:} $\ell_i \wto \ell_j$ and $\ell_j$ is a head, \ie
entering a new component (e.g., $E_1$ and $E_2$ in
Fig.~\ref{fig:nested_loops}). Then, $\tau_B(\ell_i,\ell_j)$ contains a
constraint corresponding to $c_j$ being assigned
non-deterministically.

\emph{Re-entry:} $\ell_j \wwto \ell_i$, \ie re-entering a
component via a back-edge (e.g., $R_1$ and $R_2$ in
Fig.~\ref{fig:nested_loops}). Then, $\tau_B(\ell_i,\ell_j)$ contains
the constraint $(0 \le c'_j \land c_j'=c_j-1 \land c_j \le b_j$, \ie it
decrements $c_j$ as long as it is not zero.

\emph{Exit:} $\ell_i \wto \ell_j \land \hds{\ell_i} \supset
\hds{\ell_j}$, \ie exiting (one or more) components containing
$\ell_i$ (e.g., $X_1$ and $X_2$ in Fig.~\ref{fig:nested_loops}). Then,
for each $h \in \hds{\ell_i} \setminus \hds{\ell_j}$,
$\tau_B(\ell_i,\ell_j)$ contains the constraint $\ctr(h) = 0$.

\emph{Pass-on.} For each $h \in \hds{\ell_j} \setminus
\{\ell_j\}$, $\tau_B(\ell_i,\ell_j)$ contains the constraint $\ctr(h)
= \ctr(h)'$. Thus, when the transition is inside a component the
current value of its counter is remembered. See $\tau_B$ for the
transitions $E_2$, $R_1$ and $X_1$ in Fig.~\ref{fig:nested_loops}.

In other words, a counter is assigned a non-deterministic initial
value when entering its component, and decremented until zero before
exiting.
Since the bound variables (\ie $B$) are unconstrained, $\bp{P}$ and
$P$ are equivalent w.r.t. safety, as stated by
Lemma~\ref{lem:bp_sound}.

\section{Proof Sketch of Lemma~\ref{lem:ind_invs}}
  Initially, $\invar$ is empty, denoting an invariant of $\top$ for
  every location, which is clearly inductive. The only update to
  $\invar$ is on line~31. As $\pi$ in {\sc ExtractInvs} is a \emph{safety proof}, the only invariant which can be added to
  $\ell^o$ is equivalent to $\top$. For every other location, the added invariants
  are inductive relative to $\invar$ which follows from the failure of
  the condition on line~29.  Thus, $\invar$ remains inductive.  \qed

\section{Proof of Theorem~\ref{thm:soundness}}
  If \algo returns {\sc Safe} (line~8), the condition on line~7 and
  Lemma~\ref{lem:ind_invs} imply that $\invar$ is a safety proof for
  $P$. Thus, $P$ is safe. If \algo returns {\sc Unsafe} (line~13), a
  feasible counterexample has been found (line~12). Thus, $P$ is
  unsafe.  \qed

\section{Proof of Lemma~\ref{lem:bp_sound}}
        Suppose $\bp{P}$ is unsafe and there is a feasible control path to
        $\ell^e$ and a corresponding state sequence. Now, projecting the state
        sequence from $V \cup C \cup B$ to $V$ clearly satisfies $\tau$ for
        every pair of locations as $\bp{P}$ only strengthens the transitions
with additional constraints ($\tau_B$).
        This gives us a counterexample for $P$ and $P$ is unsafe.

        Now, assume that $\bp{P}$ is safe with a safety proof $\bp{\pf}$. We
        show that $\pf : L \to \power{\bexpr{V}}$ such that for all $\ell \in L$,
        $\pf(\ell) \equiv \{\forall B \ge 0,C \ge 0 \such \varphi ~\mid~ \varphi \in \bp{\pf}(\ell)\}$ is
        a safety proof for $P$.
        Note that the quantifiers can be eliminated for Linear Rational
        Arithmetic. It suffices to show the three conditions of
        Definition~\ref{def:proof}.

        $\bigand \pf(\ell^e) \equiv \bigand_{\varphi \in \bp{\pf}(\ell^e)} \forall B \ge
0, C \ge 0 \such \varphi \equiv \forall B \ge 0, C \ge 0 \such \bigand_{\varphi \in
\bp{\pf}(\ell^e)} \varphi$. As $\bigand_{\varphi \in \bp{\pf}(\ell^e)} \varphi \implies
\bot$, $\bigand \pf(\ell^e) \implies \bot$ and the first condition is satisfied.

        As $\top \implies \bigand \bp{\pf} (\ell^o)$, $\top \implies \varphi$ for
        every $\varphi \in \bp{\pf}(\ell^o)$ and in particular, $\top \implies \forall B
        \ge 0,C \ge 0 \such \varphi$. Therefore, $\top \implies \bigand \pf(\ell^o)$ satisfying the second condition.

        Let $s,s'$ be a pair of current and next states satisfying
        $\bigand \pf(\ell_i) \land \tau(\ell_i,\ell_j)$ for some $\ell_i,\ell_j \in L$. 
        We need to prove that $\forall B \ge 0 ,C \ge 0 \such \varphi$ is true for $s'$, for every
        $\varphi \in \bp{\pf}(\ell_j)$.
        Let $b',c'$ be arbitrary non-negative values for $B,C$, respectively.
One can easily show that $\tau_B(\ell_i,\ell_j)$ is invertible for non-negative
values of the post-variables and let $b,c$ be the values of the pre-variables
corresponding to $b',c'$. But then,
        for $b,c$ and $s$, we know that $\bigand \bp{\pf}(\ell_i)$ is true. Given that
        $\bp{\pf}$ is a proof of $\bp{P}$, it follows that
        $\varphi$ is true for $b',c'$ and $s'$.
        \qed

\section{Proof of Theorem~\ref{thm:progress}}
In the following, we sometimes refer to the components of a program $P$ by
application, \eg $L(P)$, in addition to using subscripts.

\begin{lemma}
\label{lem:nextua}
Let $U_1 \under_{\sigma_1} A$ with $\tau(U_1) = \sigma_1(\tau_A) \land \rho_1$.
If $U_1 \under_\mu U_2 \under_{\sigma_2} A$ with $\sigma_1 = \sigma_2 \circ
\mu$, there exists $\rho_2$ such that $\tau(U_2) = \sigma_2(\tau_A) \land
\rho_2$ and $\rho_1 \implies \mu(\rho_2)$.
\end{lemma}

\begin{pfsketch}
Let $\tau(U_2) = \sigma_2(\tau_A) \land \rho$ (such a $\rho$ can always be
found, as $\tau(U_2) \implies \sigma_2(\tau_A)$).
As $U_1 \under_\mu U_2$, we have that $\tau(U_1) \implies \mu(\tau(U_2))$.
Together with $\sigma_1 = \sigma_2 \circ \mu$, we obtain
\[
\sigma_1(\tau_A) \land \rho_1 \implies \sigma_1(\tau_A) \land \mu(\rho).
\]
Consider
\[
\rho_2 \equiv \rho \lor \lambda \ell^2_i, \ell^2_j \such
		\left( \bigor_{\ell^1_i, \ell^1_j}
		\mu(\ell^1_i) = \ell^2_i \land
		\mu(\ell^1_j) = \ell^2_j \land
		\rho_1 (\ell^1_i, \ell^1_j) \right).
\]
It can be easily shown that $\rho_1 \implies \mu(\rho_2)$. Furthermore, it can
be shown, using $\sigma_1 = \sigma_2 \circ \mu$, that $\sigma_1(\tau_A) \land \mu(\rho) \Leftrightarrow
\sigma_1(\tau_A) \land \mu(\rho_2)$ and hence, $\sigma_2(\tau_A) \land \rho
\Leftrightarrow \sigma_2(\tau_A) \land \rho_2$.
\qed
\end{pfsketch}

\begin{figure}[t]
\centering
\begin{subfigure}[b]{.5\textwidth}
\includegraphics[scale=1]{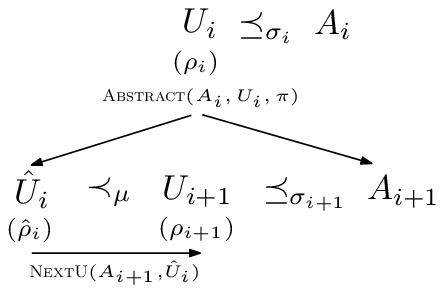}
\caption {$U_i$ is safe.}
\label {fig:progress1}
\end{subfigure}
\begin{subfigure}[b]{.4\textwidth}
\includegraphics[scale=1]{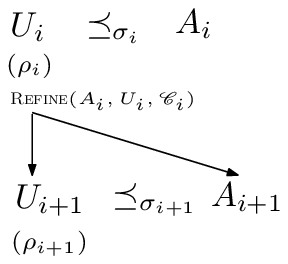}
\caption {$U_{i+1}$ is unsafe.}
\label {fig:progress2}
\end{subfigure}
\caption {Relation between two successive under-approximations $U_i$ and
$U_{i+1}$.}
\label {fig:progress}
\end{figure}

\begin{enumerate}
\item
$U_{i+1}$ is obtained from $U_i$ after a call to \textsc{Abstract} followed by
\textsc{NextU}, as shown in Fig.~\ref{fig:progress1}. For $U_j$, the figure also shows
$\rho_j$ in brackets such that $\tau(U_j) = \sigma_j(\tau(A_j)) \land \rho_j$.
\textsc{Abstract} ensures that $\rho_i \implies \hat{\rho}_i$ and
Lemma~\ref{lem:nextua} guarantees the existence of a $\rho_{i+1}$ with
$\hat{\rho}_i \implies \mu(\rho_{i+1})$. Together, $\rho_i \implies
\mu(\rho_{i+1})$. Further, \textsc{NextU} requires $\sigma_i = \sigma_{i+1}
\circ \mu$. Then, $\conc{U}_i \under_\mu \conc{U}_{i+1}$, as shown below.

\begin{align*}
\tau(\conc{U}_i) & = \sigma_i(\tau_P) \land \rho_i \\
& \implies (\sigma_{i+1} \circ \mu) (\tau_P) \land \mu (\rho_{i+1}) \\
& \implies \mu (\sigma_{i+1} (\tau_P)) \land \mu (\rho_{i+1}) \\
& \implies \mu (\tau(\conc{U}_{i+1}))
\end{align*}

To show that $\conc{U}_i \sunder \conc{U}_{i+1}$, assume for the sake of
contradiction that $\conc{U}_{i+1} \under_\omega \conc{U}_i$. Then,
as $\rho_i \implies \hat{\rho}_i$,

\begin{align*}
\tau(\conc{U}_{i+1}) &= \sigma_{i+1}(\tau_P) \land \rho_{i+1} \\
& \implies \omega (\sigma_i (\tau_P) \land \rho_i) \\
& \implies \omega (\sigma_i (\tau_P) \land \hat{\rho}_i) \\
& = \omega(\tau(\conc{\hat{U}}_i))
\end{align*}

giving us $\conc{U}_{i+1} \under_\omega \conc{\hat{U}}_i$. This contradicts
$\conc{\hat{U}}_i \sunder \conc{U}_{i+1}$ on line~16 of Fig.~\ref{fig:code}.

\item
$U_{i+1}$ is obtained from $U_i$ after a call to \textsc{Refine} as shown in
Fig.~\ref{fig:progress2}. Again, for $U_j$, the figure shows
$\rho_j$ in brackets such that $\tau(U_j) = \sigma_j(\tau(A_j)) \land \rho_j$.
\textsc{Refine} ensures that $\rho_i = \rho_{i+1}$ and $\sigma_i =
\sigma_{i+1}$. These imply that $\tau(\conc{U}_i) \Leftrightarrow \tau(\conc{U}_{i+1})$.

\item
Let $\cex_i = \langle \bar{\ell}_i, \bar{s} \rangle$.
We prove the stronger statement that for every $j > i$, there exist a control
path $\bar{\ell}_j$ in $U_j$ and a $\rho_j$ such that

\begin{enumerate}
\item $\sigma_j (\bar{\ell}_j) = \sigma_i (\bar{\ell}_i)$, and
\item $\tau(U_j) = \sigma_j (\tau(A_j)) \land \rho_j$, $\langle \bar{\ell}_j,
\bar{s} \rangle$ is feasible for the transition relation $\rho_j$ but not for
the transition relation $\sigma_j(\tau(A_j))$.
\end{enumerate}

In words, we show that the control path of $\cex_i$ is a control path in every
future $U_j$ (via $\sigma_j$) and the state sequence $\bar{s}$ is feasible when restricted
to $\rho_j$ but not when restricted to $\sigma_j (\tau(A_j))$. The latter is
sufficient to show that $U_j$ does not admit $\cex_i$.

We prove the stronger statement by induction on $j$. If $j = i+1$,
Fig.~\ref{fig:progress2} shows the relation between $U_i$ and $U_{i+1}$. Again,
\textsc{Refine} ensures that $\rho_i = \rho_{i+1}$ and $\sigma_i =
\sigma_{i+1}$. The required control path $\bar{\ell}_j$ in (a) is the same as
$\bar{\ell}_i$. Also, \textsc{Refine} ensures that $\tau(A_j)$ does not admit
$\cex_i$, satisfying (b).

Now, assume that $U_i$ satisfies (a) and (b), for an arbitrary $i$. We show that
$U_{i+1}$ also satisfies (a) and (b). If $U_{i+1}$ is obtained from $U_i$ after
a call to \textsc{Refine}, the argument is the same as for the base case above.
The other possibility is as shown in Fig.~\ref{fig:progress1} where $U_i$ is
safe and $U_{i+1}$ is obtained after a call to \textsc{Abstract}, followed by a
call to \textsc{NextU}. Consider $\rho_i$, $\hat{\rho}_i$ and $\rho_{i+1}$ as
shown in the figure.

To see that (a) is satisfied, consider the control path $\mu(\bar{\ell}_i)$ and
note that $\sigma_i = \sigma_{i+1} \circ \mu$ (line~16 of Fig.~\ref{fig:code}). 

To see that (b) is satisfied, Lemma~\ref{lem:nextua} ensures the existence of
$\rho_{i+1}$ with $\hat{\rho}_i \implies \mu(\rho_{i+1})$. As $\bar{s}$ is feasible
along $\bar{\ell}_i$ for the transition relation $\rho_i$ and hence, for $\hat{\rho}_i$, it is also
feasible along $\mu(\bar{\ell}_i)$ for the transition relation
$\mu(\rho_{i+1})$. Moreover, $\bar{s}$ is infeasible along $\bar{\ell}_i$ for
$\sigma_i(\tau(A_{i+1}))$, as $\hat{U}_i$ is safe. Hence, it remains infeasible along $\mu(\bar{\ell}_i)$ for
$\sigma_{i+1}(\tau(A_{i+1}))$ (follows from $\sigma_{i+1} \circ \mu = \sigma_i$).
\end{enumerate}
\qed


\end{document}